\documentclass[11pt, a4paper]{revtex4}
\usepackage{graphicx}
\usepackage{dcolumn}
\usepackage{latexsym}
\usepackage{amsmath, amsthm, amssymb}
\usepackage{epsfig}
\usepackage{bm}
\usepackage[dvipsnames]{xcolor}
\usepackage{caption,subcaption}
\usepackage{hyperref}
\usepackage{float}
\usepackage{aliascnt}
\usepackage{xcolor}
\newaliascnt{eqfloat}{equation}
\newfloat{eqfloat}{h}{eqflts}
\floatname{eqfloat}{Equation}

\newcommand*{\ORGeqfloat}{}
\let\ORGeqfloat\eqfloat
\def\eqfloat{%
  \let\ORIGINALcaption\caption
  \def\caption{%
    \addtocounter{equation}{-1}%
    \ORIGINALcaption
  }%
  \ORGeqfloat
}
\usepackage[utf8]{inputenc}
\DeclareUnicodeCharacter{2212}{-}

\begin{document}

\title{Inhomogeneous activity enhances density phase separation in active model B }

\author{Ajeya Krishna}
\email[]{ajeya.krishnap.phy17@iitbhu.ac.in}
\affiliation{Indian Institute of Technology (BHU), Varanasi, U.P. India - 221005}

\author{Shradha Mishra}
\email[]{smishra.phy@itbhu.ac.in}
\affiliation{Indian Institute of Technology (BHU), Varanasi, U.P. India - 221005}

\begin{abstract}
\section{Abstract}
    We study the binary phase separation in active model B, on a two-dimensional substrate with inhomogeneous activity.
	Activity was introduced with a maximum value at the center of the box and spread as a Bivariate-Gaussian distribution as we move away from the center.
	The system was studied for three different intensities of the distribution.
	Towards the boundary of the box, activity is zero or the model is similar to the passive model B. We start from
	the random homogeneous distribution of density of particles, and the system evolves towards a structured distribution of density. With time, density 
	starts to phase separate with maximum density at the center of the box and decreases as we move away from the center of the box.   
	The width of the density profile at the center increases as a  power law exponent $\alpha(t)$ remains close between 2/3 to 3/4 up to some moderate time and then decays to zero in the steady state. 
	Hence, our result shows the response of density in an active binary system with respect to
	the patterned substrate. It  can be used to design devices useful for the trapping and segregation of  active particles.

	%a collection of active Brownian particles, using coarse-grained hydrodynamic equation of motion for local active model B, with inhomogeneous activity ase separation among disoriented active particles whose speed decreases swiftly enough with order-parameter which is density in this model. We change the constant distribution of activity into a Gaussian distribution from the center of the box. This is scalar $\phi^4$ scalar field theory which includes the activity. The activity is the square gradient term, that breaks the time-reversal symmetry of the model. We observe that this change in activity distribution has unpresuming effects on coarsening dynamics. The accumulation of active particles happens unusually by creating a cluster of high density particles(A-atoms) at the center (where the activity is highest). In the  locality where there is no activity, atoms follows primary phase separation mechanisms of passive particles. The results are quite surprising and are yet to be explained statistically.
\end{abstract}

\maketitle
\section{Introduction}
Ranging from small organisms like bacteria \cite{bacteria}, algae to higher organisms like fish \cite{fish}, birds \cite{bird}, animals self-organize 
themselves to form complex structures \cite{marchetti}. This topic has been under active research for the past decade \cite{pna}. 
The emergence of meso-scale\cite{meso1,meso2,meso3,meso4} turbulent motion was a forward step in research on non-equilibrium biological systems.

Consider the phase separation kinetics of a binary (AB) mixture, which has a conserved order parameter. The initially homogeneous system divides into A-rich and B-rich domains. In contrast to the non-conserved case, the evolution in this case must abide by the constraint that the numbers of A and B remain constant, i.e. the order parameter must be conserved. Natural examples include oil and vinegar phase separation. Initially, the oil and vinegar are combined (AB mixture). This homogeneous structure eventually divides into oil-rich (A-rich) and vinegar-rich (B-rich) zones.
Particles in the preceding case are passive, i.e., they do not move on their own, but rather as a result of the effect of external parameters. Since we were more interested in particles that propelled themselves, we included activity in Passive Model B. This model is known as Active Model B and the activity was consistent in the system. We considered integrating inhomogeneous activity and studying the resulting process. As a consequence, we included inhomogeneous activity.

%Self-propulsion and their mutual interactions drives the microorganisms which simply referred as active particles (self propulsion) \cite{bacteria}. 
%Systems with medium or long-range orientation order of these active systems are successfully described by adding minimal active terms to equation of hydrodynamics for liquid crystals. 
%The resulting mean field theories involves either vector or tensor order parameter describing their phase separation mechanisms.\cite{rama,marchetti}. 
Dynamics of active colloidal particles such as natural microorganisms like bacteria or algae \cite{marchetti,cates}, or synthetic swimmers, active Brownian particles (ABP)
\cite{howse,ebbens,thutupalli,volpe,palacci} described by  having a non-trivial dependence of active current to the local curvature 
of the underlying density profile.  This leads to an activity term in standard binary phase separation in equilibrium system or also 
called as passive model B \cite{chc2,chc,chc3}. The corresponding active model is called the active model B (AMB) \cite{ncomms5351}. 
%This type of
%and active term absent 
The study of passive model B \cite{chc2,chc,chc3} is useful to understand the phase separation in equilibrium binary systems \cite{chc2,chc}, whereas active model B, 
gives the understanding of phase separation in many natural and biological systems. Also, many artificially designed active Janus particles in the lab are 
also useful candidates for technological and pharmaceutical applications \cite{pattanayak,malgaretti,bao}. 
%is studied in detail Study of phase separation of active particles system is a very interesting, challenging research in the field of Soft Matter Physics. 
Active Brownian particles predominantly show short-range steric repulsion \cite{smishra1}. The presence of activity shows fascinating behavior like coherent motion, 
phase separation without any external parameter or quenching in temperature \cite{rama,marchetti}. 
%Generally phase separation mechanisms of Active particle system is 
%described by taking vector or tensor order parameter.
The phase separation of systems with self-propelled particles (active particles) is studied numerically 
\cite{smishra1,smishra2,fily,tiribocchi} and to some extent in experimental studies \cite{butti,liu}.

In this study, we consider a conserved scalar order parameter $\phi$ which is the local density of the particles.
The motivation of selecting a scalar order parameter is Model B \cite{ncomms5351, puri}. The resulting phase-separation kinetics of passive 
colloidal particles is best described by a conserved scalar order parameter $\phi$ (continuous local density parameter). 
Simplifying the free energy to quadratic polynomial in $\phi$ using general diffusion mechanisms, gives a theory of ``Model B''. 
Model B or $\phi^4$ field theories are the simplest form of the Cahn-Hilliard equation \cite{chc2,chc,chc3}. 
%The $L \alpha t^\frac{1}{3}$ power law is the result of this findings.

The kinetics of phase separation of passive and active systems are extremely distinct but coarse-graining of active systems at a large scale demonstrates some relation between the two cases. 
The relation was first observed in models of swimming bacteria with discrete reorientations \cite{tai} and later extended to ABPs \cite{cate}.

%We consider a system with mixture of two  species A and B with $\phi(r)$ which represents average value of the excess concentration of A-atoms over a spatial cell.
%The scalar $\phi(r)$ defined to lie on $[-1,1]$. The extremum values, $-1$ represents an excess of B-atoms and $+1$ represents an excess of A-atoms\cite{chc2}.
%In this model, we are studying such model. 
In our model, we are adding a non-integrable gradient term to passive Model B \cite{chc2,chc,chc3}. The added gradient term simply breaks detailed balance in Passive Model B, 
which suggest that active model B cannot be derived from a free-energy functional \cite{ncomms5351}. 
In this active model, we add $\lambda|\nabla\phi|^2$ term to the derivative of free-energy functional or simply chemical potential. 
The $\lambda$ here determines the strength of activity in the system. In previous studies, \cite{ncomms5351,rakesh}, the activity $\lambda$ is considered a constant value or uniform over the system. In nature, bacteria, for example, may come across an activity source such as food or some other source that increases their speed. It's more likely that the source isn't uniform. To mimic the natural activity source, we decided to use a Gaussian distribution. We thought up the bivariate normal distribution, which is equally distributed in both the x and y directions. The results could help advance technology.
In this work, we consider $\lambda$ inhomogeneous in space and take it as  Bivariate-Gaussian distributed in the system. We consider a two-dimensional square box with 
periodic boundary conditions, where the activity is chosen maximum at the center of the box and decay as a Bivariate-Gaussian distribution as we go away from the center. 
%The intensity at the width 
%of the distribution is varyied, in such a manner that volume inside distribution remains constant. 
The system is studied for three
different values of maximum intensity at the center. We call the model the inhomogeneous active model B (IAMB). For comparison, we also studied the passive model B (PMB) and constant activity model (AMB), where $\lambda$ remains constant in the whole system. 
%The Gaussian Distribution is deployed by considering 'Multi variate Normal Distribution' with mean $(\mu)$ at the center of the system, standard deviation $(\sigma)$ as $\frac{1}{4}$*(L). 
%We have taken three different intensities to study the effect of varying activity. After altering the type of distribution, activity parameter ($\lambda$) is now varying with space within in the system.  
Below we report the results of  steady state and kinetics of IAMB. 
%We consider $3$ different intensities of the distribution and report the
%respective time evaluation of phase separation kinetics and growth of length at the center of the box. 
We observed that
 increasing activity at the center of the box, leads to trapping of the particles at the center, hence the higher the activity more the value of local density. We later studied the growth kinetics of density in the middle of the box for three different intensities and find 
the density grow as a power law with time.
%\bf this will move in method section of calculation of L(t), We choose the middle as our starting point for the length calculation because the mean of the distribution is located there. At any given point in time, the length is the average of the lengths of accumulation in the x and y directions. This was done for various occasions, so the length is derived.
The power law exponent $\alpha(t)$ remains close between $2/3$ to $3/4$ up to some moderate time and then decays to zero
in the steady state.
%as we go from low to high intensity in the system.along with correlation plots and Length and time comparison plot. 
%We find that altering the activity parameter greatly effects the dynamics of the system.

The rest of the article is divided in the following manner. In next section \ref{model}, we first describe our  model then in section \ref{results},  we discuss our results in detail and finally conclude in section \ref{discussion}.

\section{Model and numerical details}
\label{model}

%With the principles outlined in the above section, we adopt the following dynamics of
We consider a conserved scalar order parameter field $\phi(\bf{r},t)$ at position {$\bf r$} and time $t$ in two dimensions. The variable $\phi$ is the local number density.
%The variable $\phi$ is linearly related to the local number density $\rho(r,t)$of active particles by the transformation $\phi=\frac{(2\rho-\rho_H-\rho_L)}{(\rho_H-\rho_L)}$, where $\rho_H$ and $\rho_L$ are the  high and low local densities co-existing phases respectively.
The dynamical equation for the rate of change of $\phi$ is given by continuity equation \cite{fick,fick2}:

\begin{equation}
	\frac{\partial \phi}{\partial t}= -\nabla\cdot \bf{J}
    \label{first_eq}
\end{equation}
\begin{equation}
    \bf{J} = -\nabla \mu 
    \label{second_eq}
\end{equation}
\begin{equation}
    \mu = -\phi+\phi^3-\nabla^2\phi + \lambda|\nabla\phi|^2 
    \label{third_eq}
\end{equation}

%In Multi-variate distribution,  x is $n$ dimensional vector, $x=(x_1,x_2,x_3,....,x_n)$, m is $n$-dimensional mean vector and $\Sigma$ is $n \times n$ co-variance matrix.
The  expression in eq. \ref{first_eq} represents the conservation of $\phi$ and the expression in eq. \ref{second_eq} expresses the relation between the mean current ${\bf J}$ and the 
non-equilibrium chemical potential $\mu$. The mean current ${\bf J}$ is proportional to the negative gradient of the chemical potential $\mu$. 
%The vector $\Lambda$ refers to Gaussian White Noise whose variance is taken as constant.  Variance can be multiplicative as well \cite{shradhasudiptapuri}, but taking  constant 
%variance is standard practice in active and passive Model B. This Gaussian White Noise does not  affect the phase separation kinetics in passive model  B
%and hence we  neglect the noise in our study.

%The above expression can be derived by using Fick's First and Second Laws of Diffusion \cite{fick,fick2}. Using Fick's first law, we can derive diffusion equation.
%\begin{equation}
%    J=-D \nabla \phi
%    \label{firstlaw}
%\end{equation}
%where $J$ is current or flux of atoms and $D$ is diffusion constant.
%Using continuity equation,
%\begin{equation}
%    \centerline{$\frac{\partial\phi}{\partial t}+\nabla \cdot J=0$}
%    \label{cont}
%\end{equation}
%%To achieve equilibrium in this system, the chemical potential should be constant and the total density in binary mixture or in our system remains constant. 
%The current or flux equation can be written as a gradient  of the chemical potential,
%\begin{equation}
%     J = -M \nabla \mu
%\end{equation}
%$M$ is the mobility tensor,
%\vspace{\baselineskip}

The chemical potential $\mu = \mu_0 + \mu_1$ is the sum of bulk and gradient contributions. In equilibrium, $\mu$ can be obtained by the variation of the free energy functional. 
The bulk part $\mu_0$ can be obtained from the polynomial terms in the standard  $\phi^4$ free energy functional, $\mu_0=\frac{d}{d \phi}f_0$, where  $f_0 = -\frac{\phi^2}{2}+\frac{\phi^4}{4}$. 
Hence, $\mu_0= -\phi+\phi^3$.

The gradient term $\mu_1$ has two terms, $\mu_1=\mu_1^p+\mu_1^a$. The passive gradient term is $\mu_1^p=-\nabla^2\phi$, which can be obtained by
variation of gradient term in Landau-Ginzburg free energy functional \cite{book}.
The $\mu_1^a$ is the simplest addition to chemical potential.
The  active gradient term is the same as obtained for  active model B,  $\mu_1^a = \lambda|\nabla\phi|^2$ \cite{ncomms5351}. The activity gradient term is non integrable, this is related to the fact that this cannot be derived from free energy functional.

%In our present study it is {\em inhomogeneous}  $\lambda$, instead of constant value. 
We write $\lambda$ as Bivariate-Gaussian Distribution \cite{lamda} which is given as eq. \ref{eq:lamda} ,

\begin{equation}
    \lambda(x,y) = \lambda_0 \times \frac{1}{2\pi\sigma_X\sigma_Y\sqrt{1-\rho^2} }\exp(-\frac{1}{2(1-\rho^2)}[(\frac{x-\mu_X}{\sigma_X})^2 -\frac{2\rho(x-\mu_X)(y-\mu_Y)}{\sigma_X \sigma_Y} + (\frac{y-\mu_Y}{\sigma_Y})^2])
    \label{eq:lamda}
\end{equation}
\vspace{0.5cm}

In eq. \ref{eq:lamda}, $\mu_X,\mu_Y$ are the mean along $x$ and $y$ directions respectively. In our case, it is the center of the system. $\sigma_X$ = $\sigma_Y$ = $\sigma$ represents variance along $x$ and $y$ directions. 
In our case it is $\frac{1}{4}$ * (size of system).
In eq. $\ref{eq:lamda}$, $\rho$ refers to the correlation of distribution in $x$ and $y$. In our case, there is no correlation  i.e., $\rho =  0$.
$x$ and $y$ in eq. \ref{eq:lamda}, represents x and y coordinates of the system. $\lambda_0$ determines the maximum intensity of the distribution.\\
We studied the model for three 
different cases $G= [\lambda_0, \sigma]$: $G=[1.0,L/4]$, $G=[2.0,L/4]$, $G=[3.0,L/4]$, where the first number in the square bracket represents
the maximum intensity at the center and the second  number denote the spread or variance of the distribution. 
%intensities are varied such that 
%total activity remains constant on the surface. 

%In Fig., \ref{lamda} we have graphically visualized the Bivariate-Gaussian distribution in color plot.

%{\bf Intensity plots for diferent $\lambda_0$ }:- Here, we graphically represent the different intensities of Gaussian distribution of activity. 
%We included 2D plots in Fig. \ref{lamda}, in which center peak has highest value and value decreases confirming the  Gaussian distribution.
% Distribution

%\begin{figure}[H]
   % \centering
%\begin{subfigure}{0.25\textwidth}
  
  %\includegraphics[width=\linewidth]{updated_2d_lamda_1.pdf}
  %\caption{}
  %\label{fig:1_lamda}
%\end{subfigure}\hfil % <-- added
%\begin{subfigure}{0.25\textwidth}
  
  %\includegraphics[width=\linewidth]{updated_2d_lamda_2.pdf}
  %\caption{}
  %\label{fig:2_lamda}
%\end{subfigure}\hfil % <-- added
%\begin{subfigure}{0.25\textwidth}
  
  %\includegraphics[width=\linewidth]{updated_2d_lamda_3.pdf}
  %\caption{}
  %\label{fig:3_lamda}
%\end{subfigure}
%\caption{ Graphical representation of the distribution %$\lambda(x,y)$. (a) $\lambda_0=1$, i.e., $G=[1.0,32]$. (b) %$\lambda_0=2$, i.e., $G=[2.0,32]$.
%(c) $\lambda_0=3$, i.e., $G=[3.0,32]$. Color bar represents the %local intensity of activity. }
%\label{lamda}
%\end{figure}

{\bf Numerical Details:} We have performed numerical analysis on our model using numerical methods to solve 
differential eqs. \ref{first_eq}, \ref{second_eq}, \ref{third_eq}. We randomly initialized $\phi$ and calculated 
chemical potential ($\mu$) which is given as in eq. \ref{third_eq}. Computing chemical potential ($\mu$), we can 
calculate flux(${\bf J}$) which is given as in eq. \ref{second_eq}. Then the $\phi$ is updated using flux (${\bf J}$) from 
eq. \ref{second_eq} and inserting it in eq. \ref{first_eq}. The stochastic differential equations of this model 
were solved by using Euler's numerical method \cite{euler}. The essential parameters we have considered are, $dx=1.0$, $dy=1.0$, $dt=0.01$. 
We considered a critical system i.e., number of A-particles equals the number of B-particles. 
The simulation time for the system was 5000 and the actual time was 50. The scalar order parameter $\phi(r,t)$ was randomly initialized,  
with the highest value as $0.5$ and the lowest value as $-0.5$. The mean and variance of this distribution were $0.0$ and $0.5$ 
respectively. The system size was $256 \times 256$. Periodic Boundary Conditions(PBC) were used in this model.
%with a view 
%to confine the particles inside the boundaries of system. 
In the case of Bivariate-Gaussian distribution of $\lambda$, the essential 
parameters are $\mu_X$, $\mu_Y$, $\sigma$, and $\lambda_0$. Parameters $\mu_X$ and $\mu_Y$ determines the mean 
of the distribution. In this model, the center of the system is considered as origin, so $\mu_X=0$ and $\mu_Y=0$. Parameters 
$\sigma$  determines the variance of our model. In our model, variance is $\frac{1}{4}$*(size of system). So variance along $x$ and $y$ directions  $\sigma=L/4$. 
Parameter $\lambda_0$ determines the intensity of distribution. The length is measured for $\lambda_0=1,2,3$ intensities. To calculate length, we choose the middle of the box as our starting point for the length calculation because the mean of the distribution is located there. At any given point in time, the length is the average of the lengths of accumulation of density above the mean value in the x and y directions. This was done for various occasions, hence  length is calculated.

\section{Results}
\label{results}

Now we discuss our results in detail: 
We integrate the nonlinear partial differential equation for local density with mean density $\phi_0=0.0$ for three different cases $G=[1.0,L/4]$, $G=[2.0,L/4]$, $G=[3.0,L/4]$.
%We have included simulations of different variations of the model we assum
Firs, we consider passive model B and active model B with constant activity parameter (AMB). After that, we consider the three different intensities of Gaussian distribution in the model.

{\bf Passive model B (PMB)}:- In this section, we will discuss the results of passive model B. This model doesn't have any activity term and the box size is $L=256$. This study is performed to understand the comparison between the passive model and Bivariate Distribution of activity parameter (IAMB) We considered a critical mixture for this study. We observe a clear difference in the kinetics of domain growth between the two cases. In this case, as time progress, unlike accumulation in the IAMB model, we see A-particles and B-particles forming connected domains. The simulation time is 50000 and the actual time is 500.

\begin{figure}[H]
    \centering % <-- added
\begin{subfigure}{0.25\textwidth}
  \includegraphics[width=\linewidth]{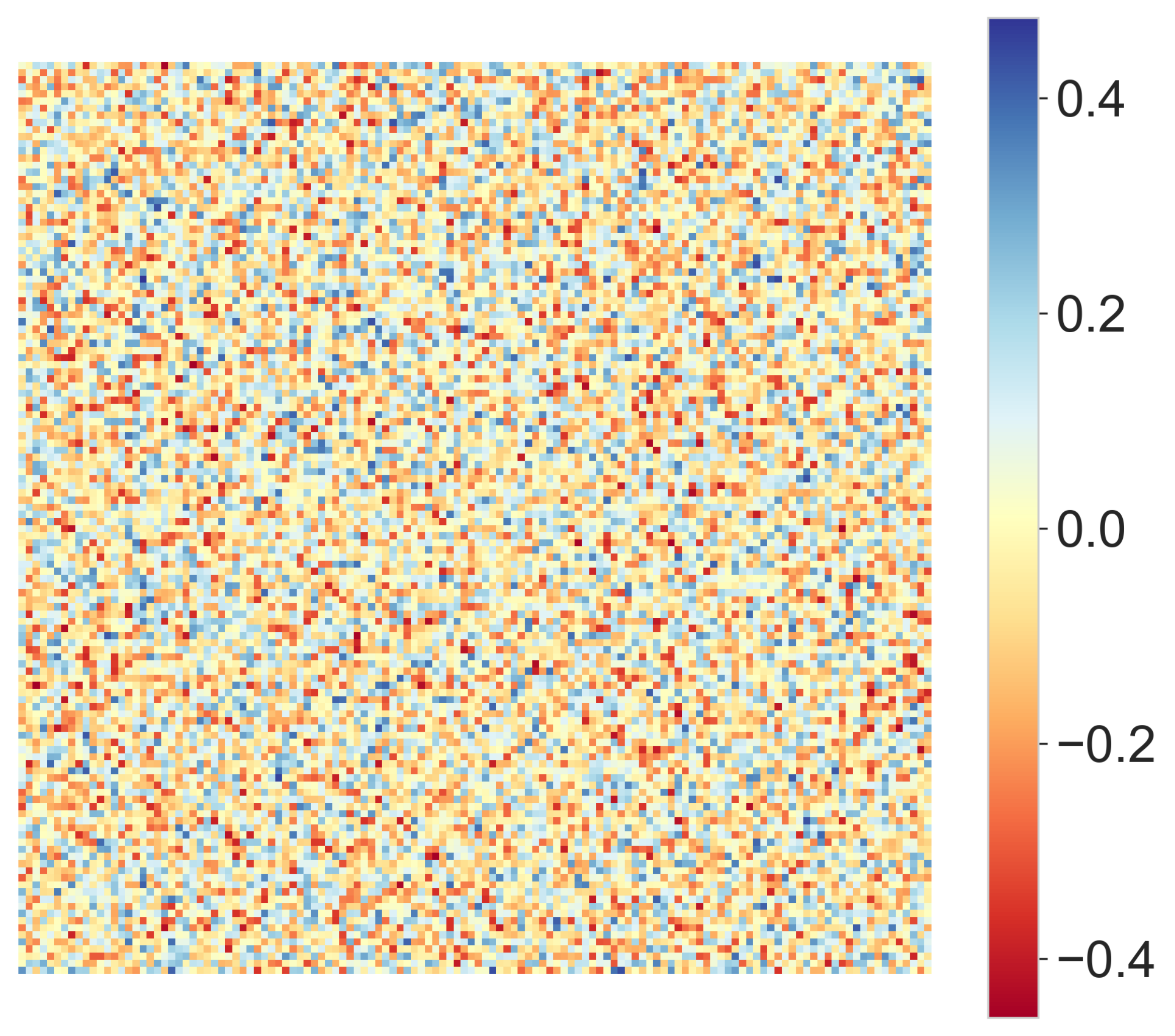}
  \caption{}
  \label{fig:1_pas}
\end{subfigure}\hfil % <-- added
\begin{subfigure}{0.25\textwidth}
  \includegraphics[width=\linewidth]{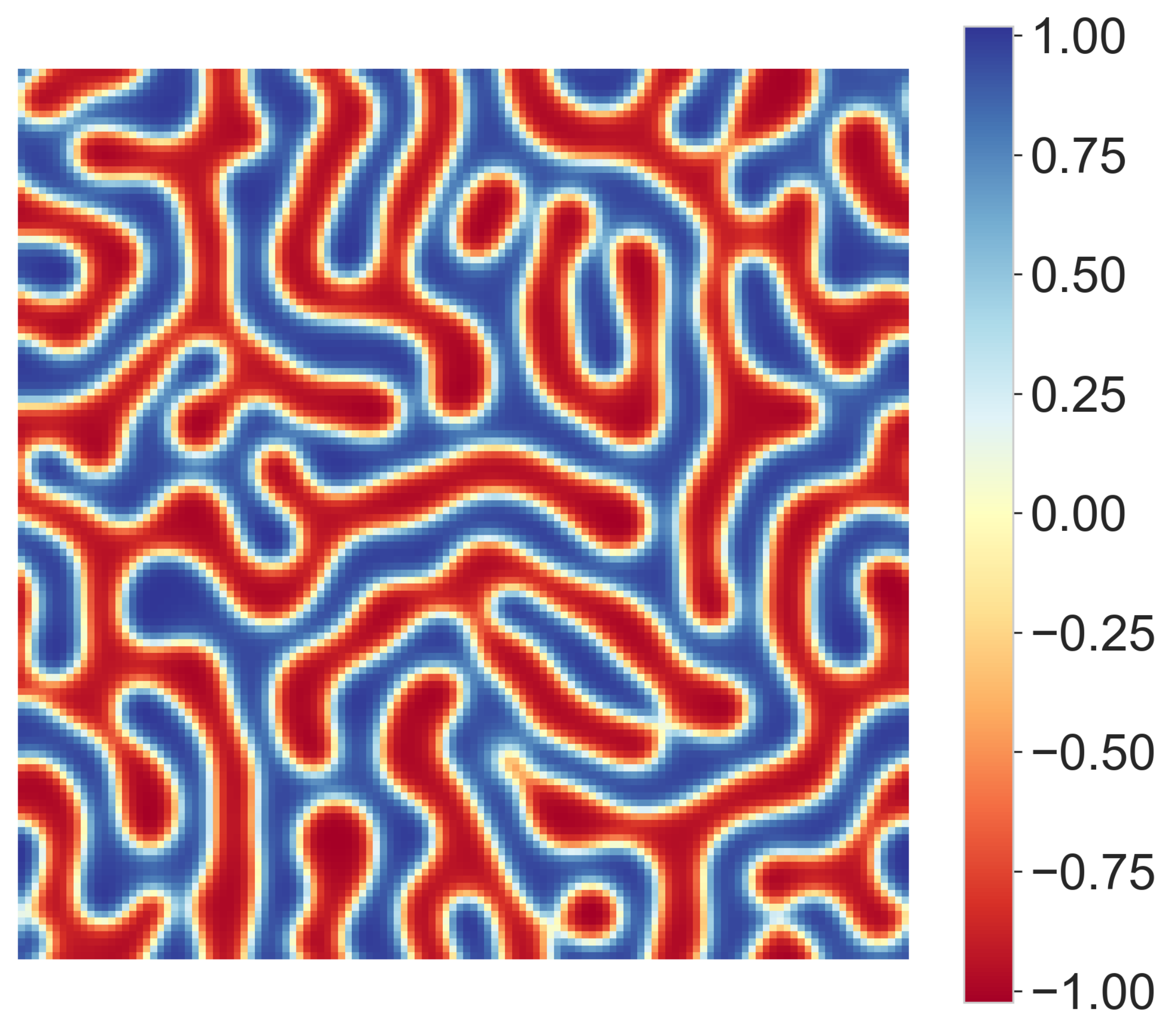}
  \caption{}
  \label{fig:2_pas}
\end{subfigure}\hfil % <-- added
\begin{subfigure}{0.25\textwidth}
  \includegraphics[width=\linewidth]{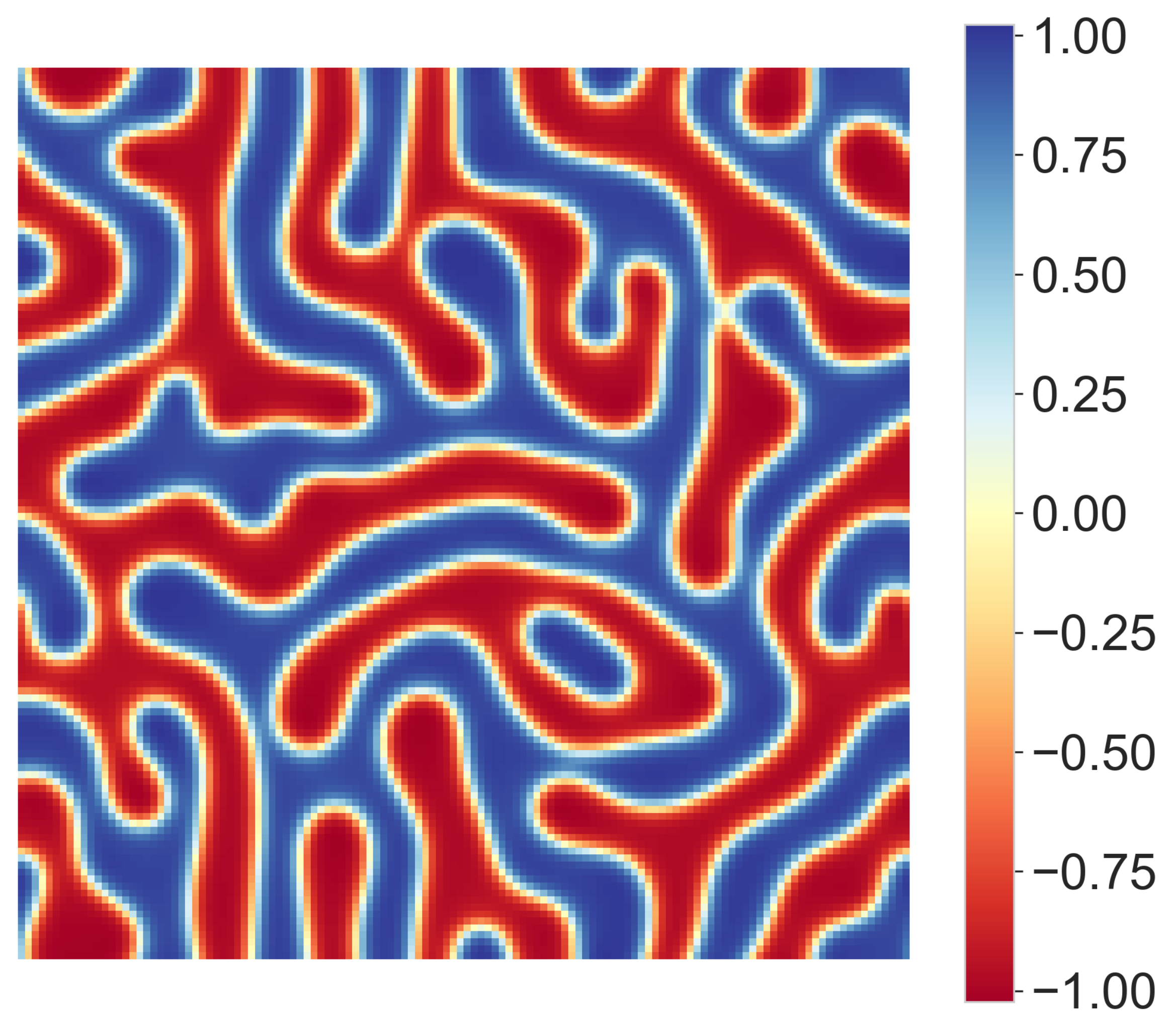}
  \caption{}
  \label{fig:3_pas}
\end{subfigure}

\medskip
\begin{subfigure}{0.25\textwidth}
  \includegraphics[width=\linewidth]{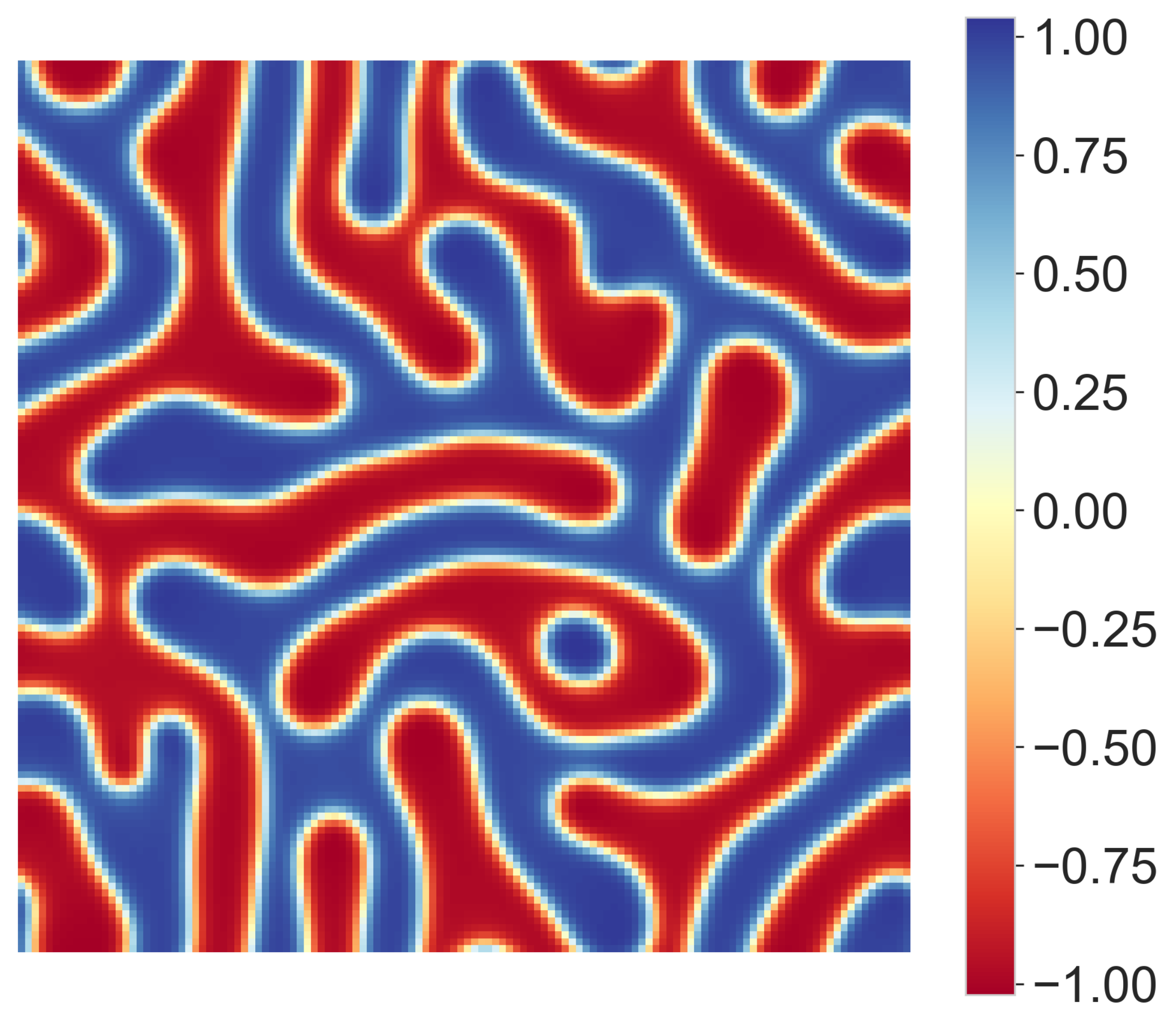}
  \caption{}
  \label{fig:4_pas}
\end{subfigure}\hfil % <-- added
\begin{subfigure}{0.25\textwidth}
  \includegraphics[width=\linewidth]{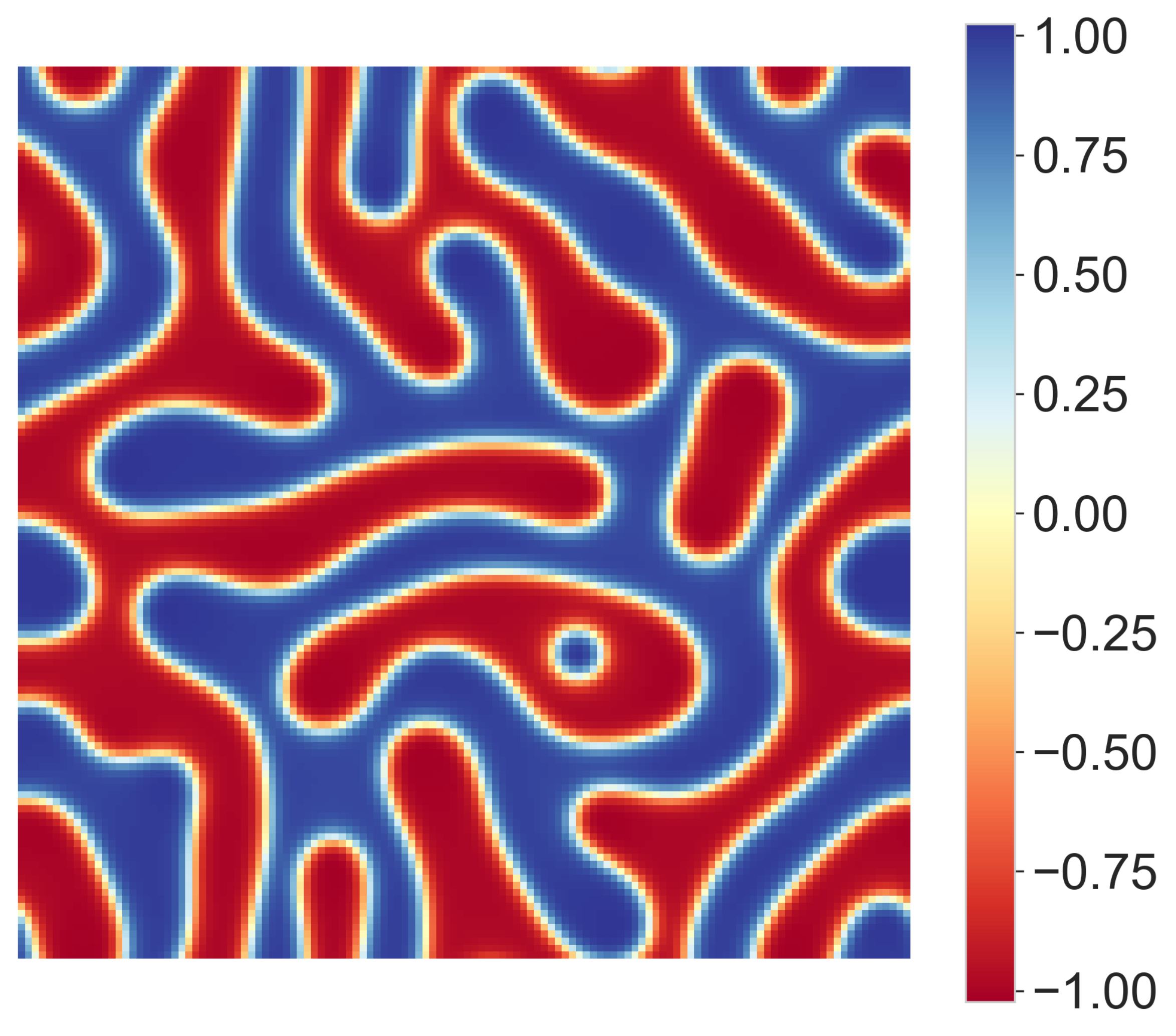}
  \caption{}
  \label{fig:5_pas}
\end{subfigure}\hfil % <-- added
\begin{subfigure}{0.25\textwidth}
  \includegraphics[width=\linewidth]{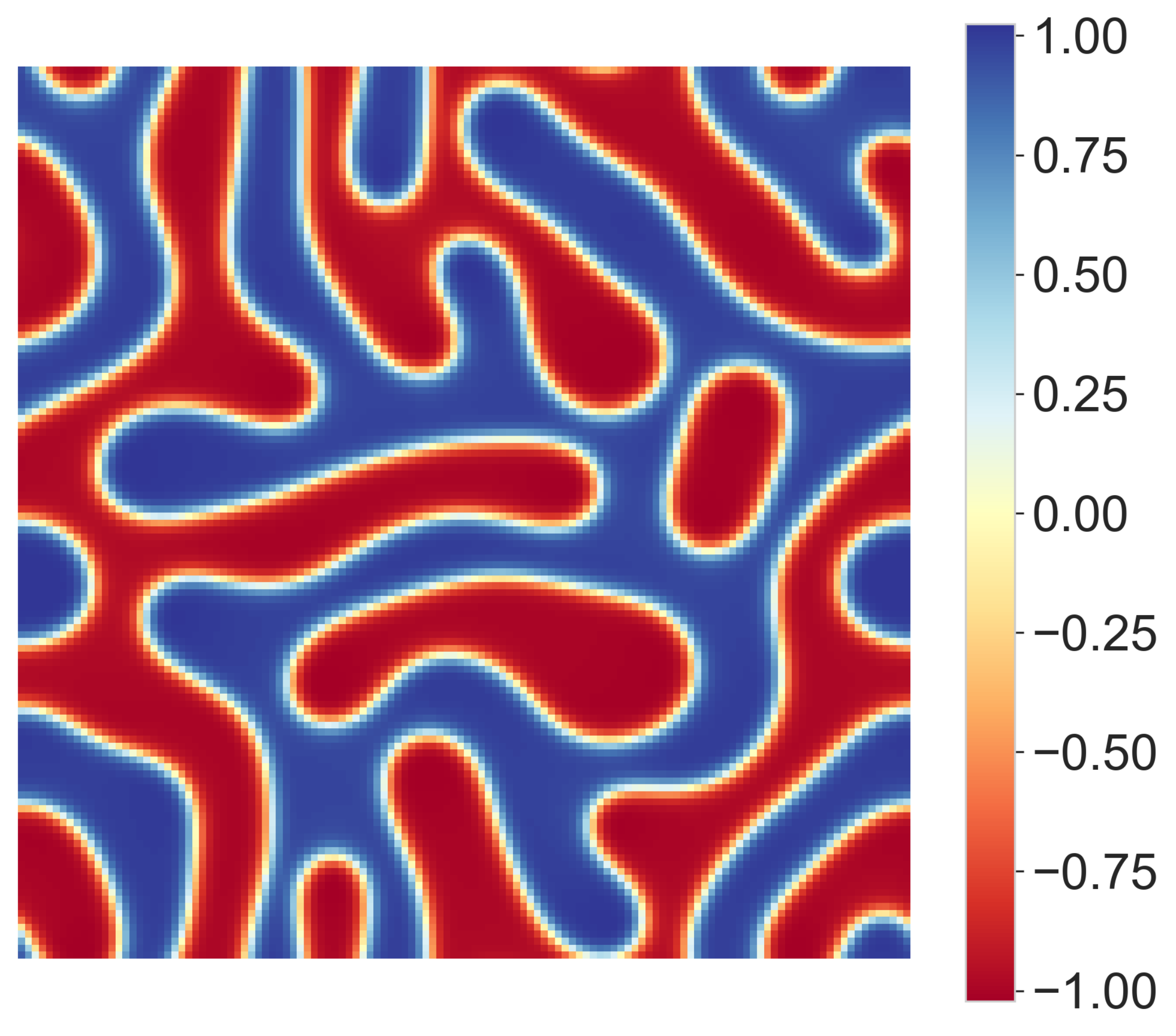}
  \caption{}
  \label{fig:6_pas}
\end{subfigure}
	\caption{Density evolution snapshots of the Passive model B. (a) at t=0, (b) at t=100, (c) at t=200, (d) at t=300, (e) at t=400, (f) at t=500. The color bar represents the value of local density $\phi$.}
\label{fig:passive}
\end{figure}

\begin{figure}[H]
    \centering % <-- added
\begin{subfigure}{0.25\textwidth}
  \includegraphics[width=\linewidth]{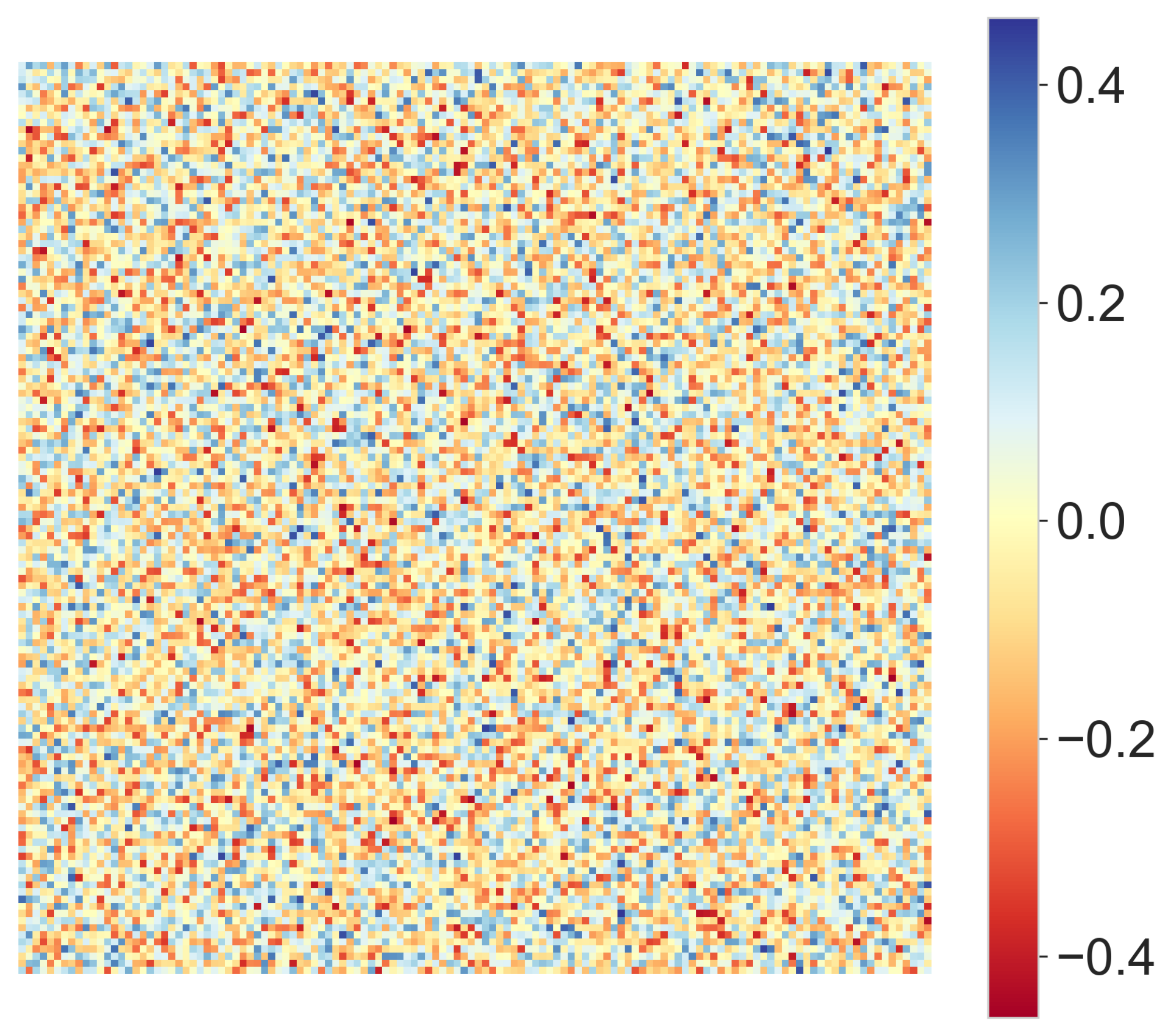}
  \caption{}
  \label{fig:1_act}
\end{subfigure}\hfil % <-- added
\begin{subfigure}{0.25\textwidth}
  \includegraphics[width=\linewidth]{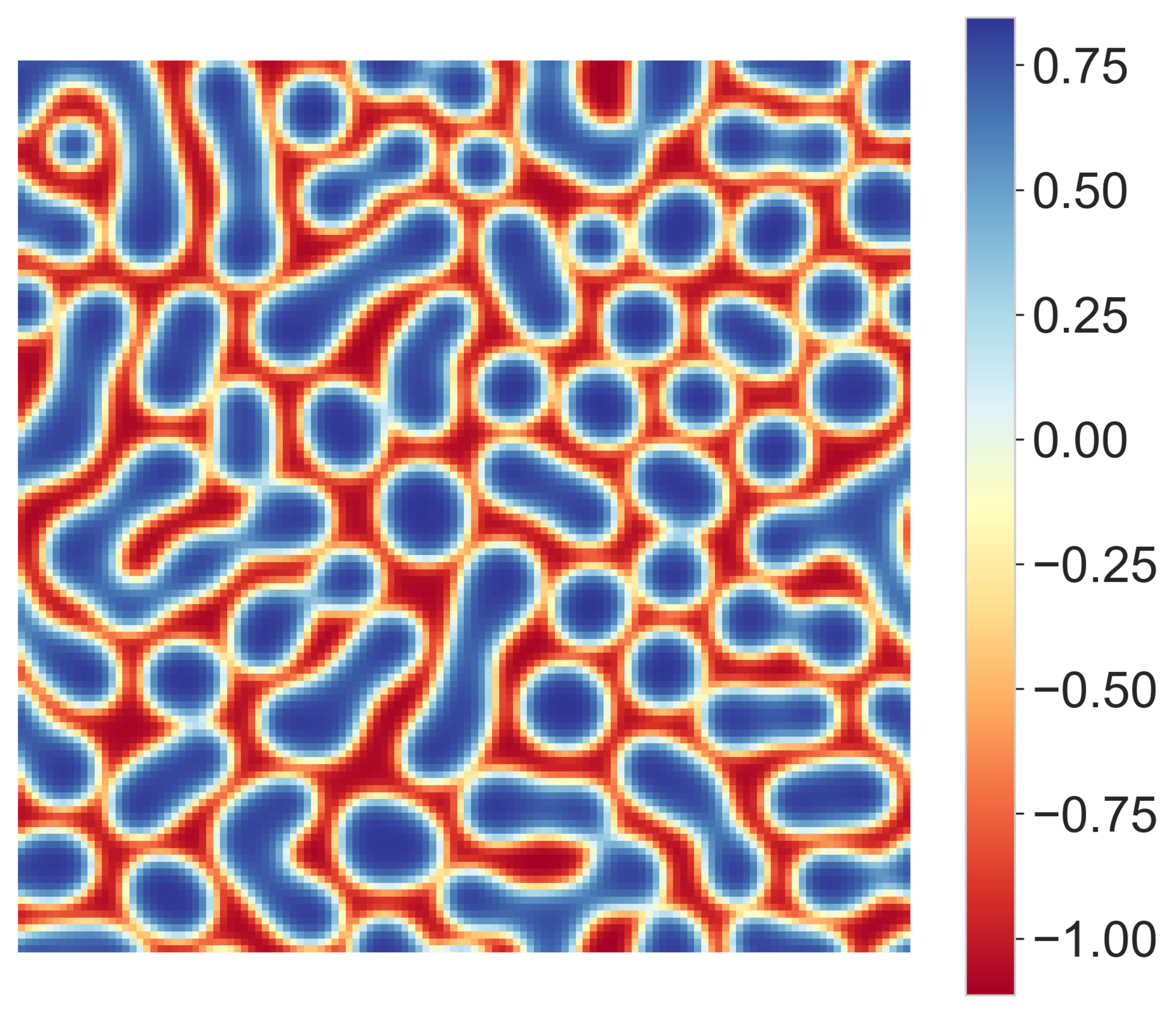}
  \caption{}
  \label{fig:2_act}
\end{subfigure}\hfil % <-- added
\begin{subfigure}{0.25\textwidth}
  \includegraphics[width=\linewidth]{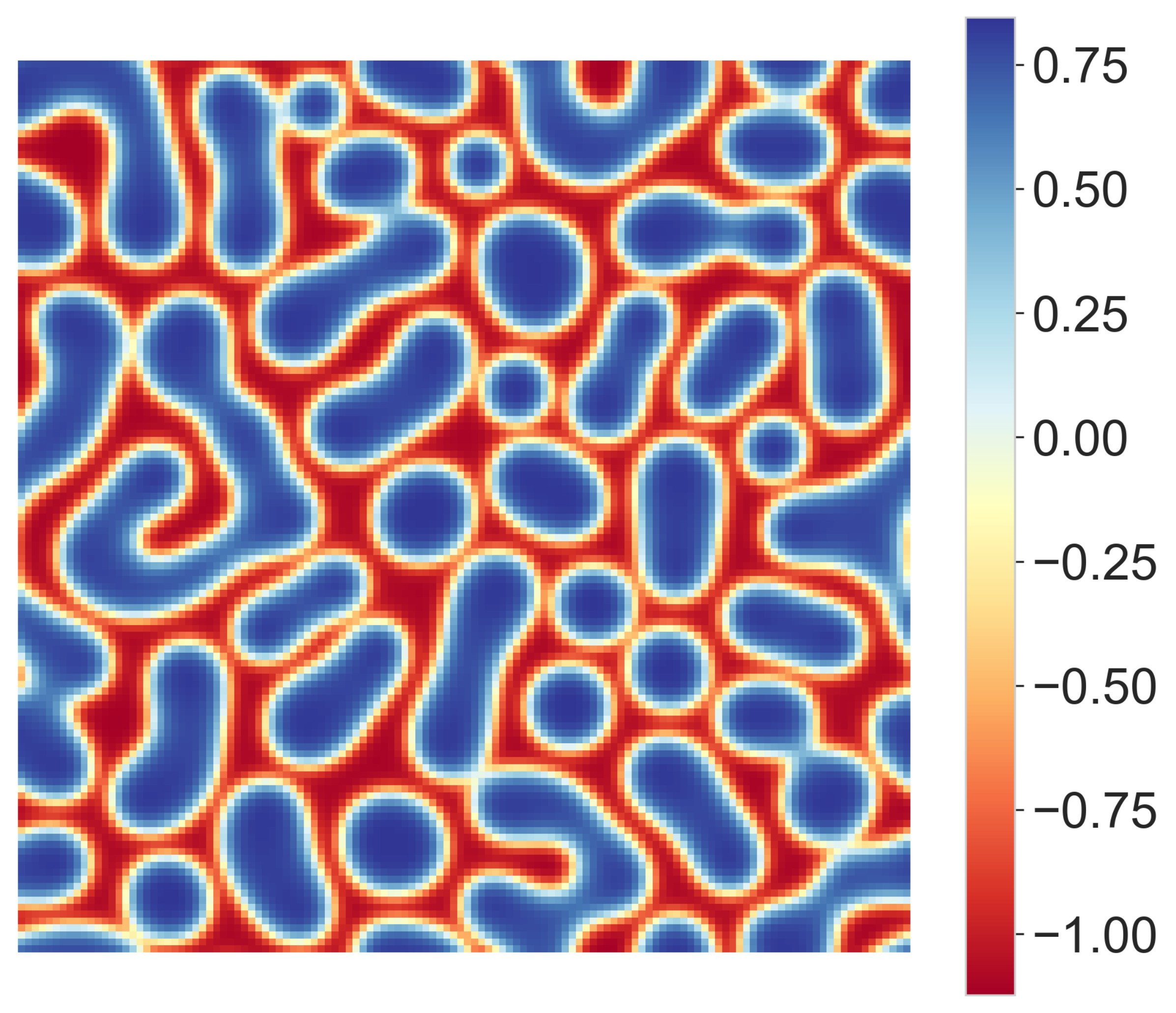}
  \caption{}
  \label{fig:3_act}
\end{subfigure}

\medskip
\begin{subfigure}{0.25\textwidth}
  \includegraphics[width=\linewidth]{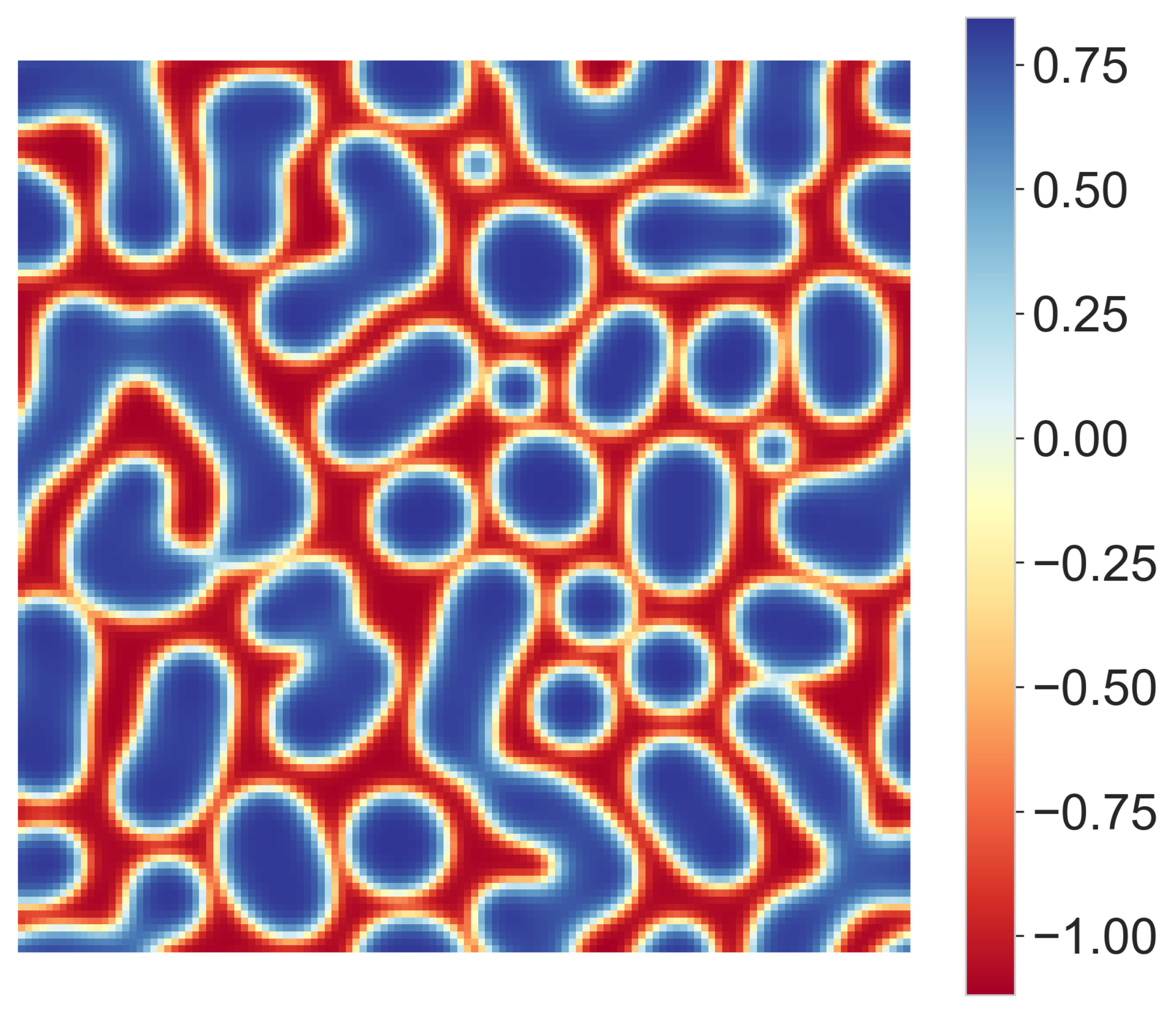}
  \caption{}
  \label{fig:4_act}
\end{subfigure}\hfil % <-- added
\begin{subfigure}{0.25\textwidth}
  \includegraphics[width=\linewidth]{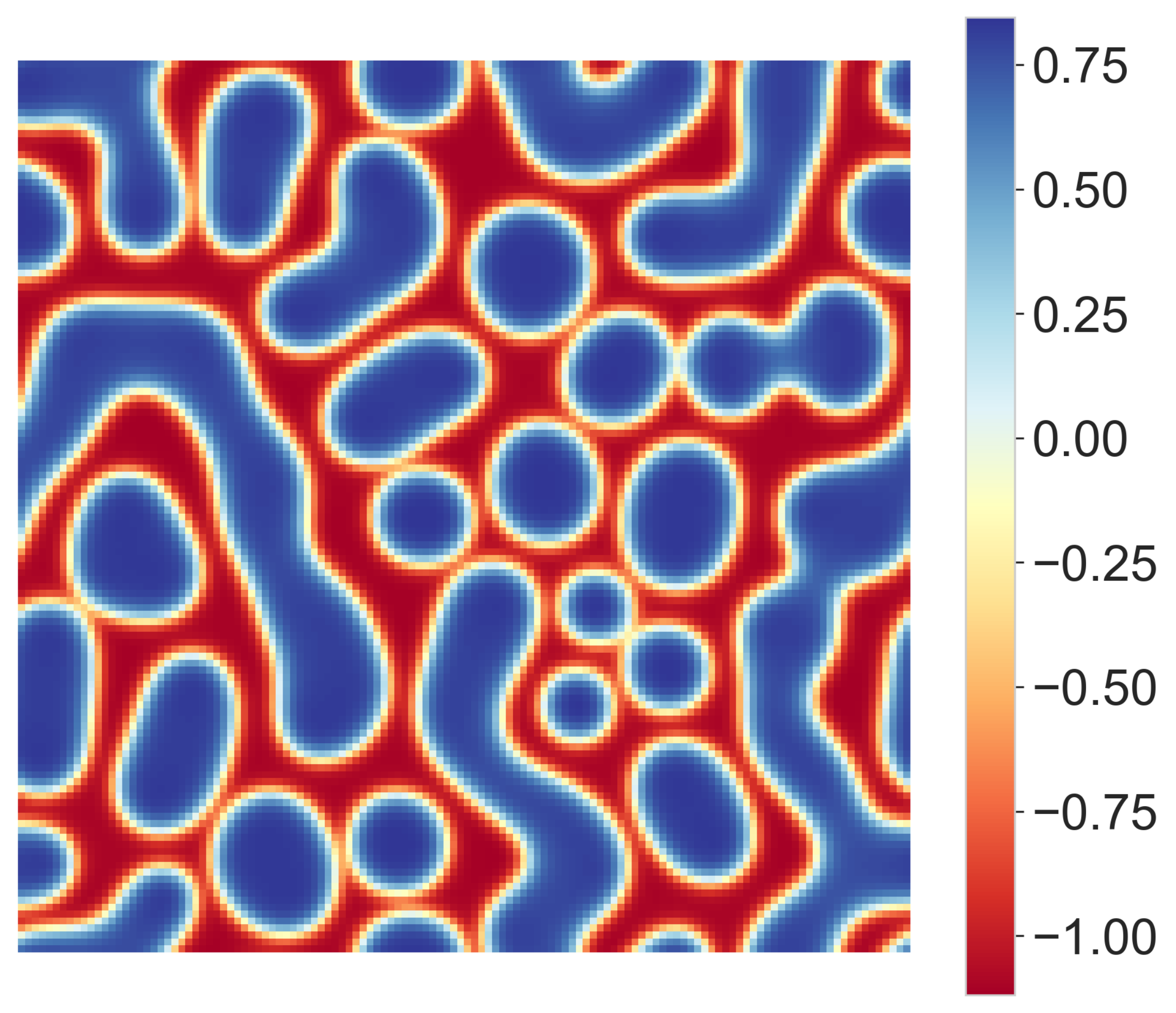}
  \caption{}
  \label{fig:5_act}
\end{subfigure}\hfil % <-- added
\begin{subfigure}{0.25\textwidth}
  \includegraphics[width=\linewidth]{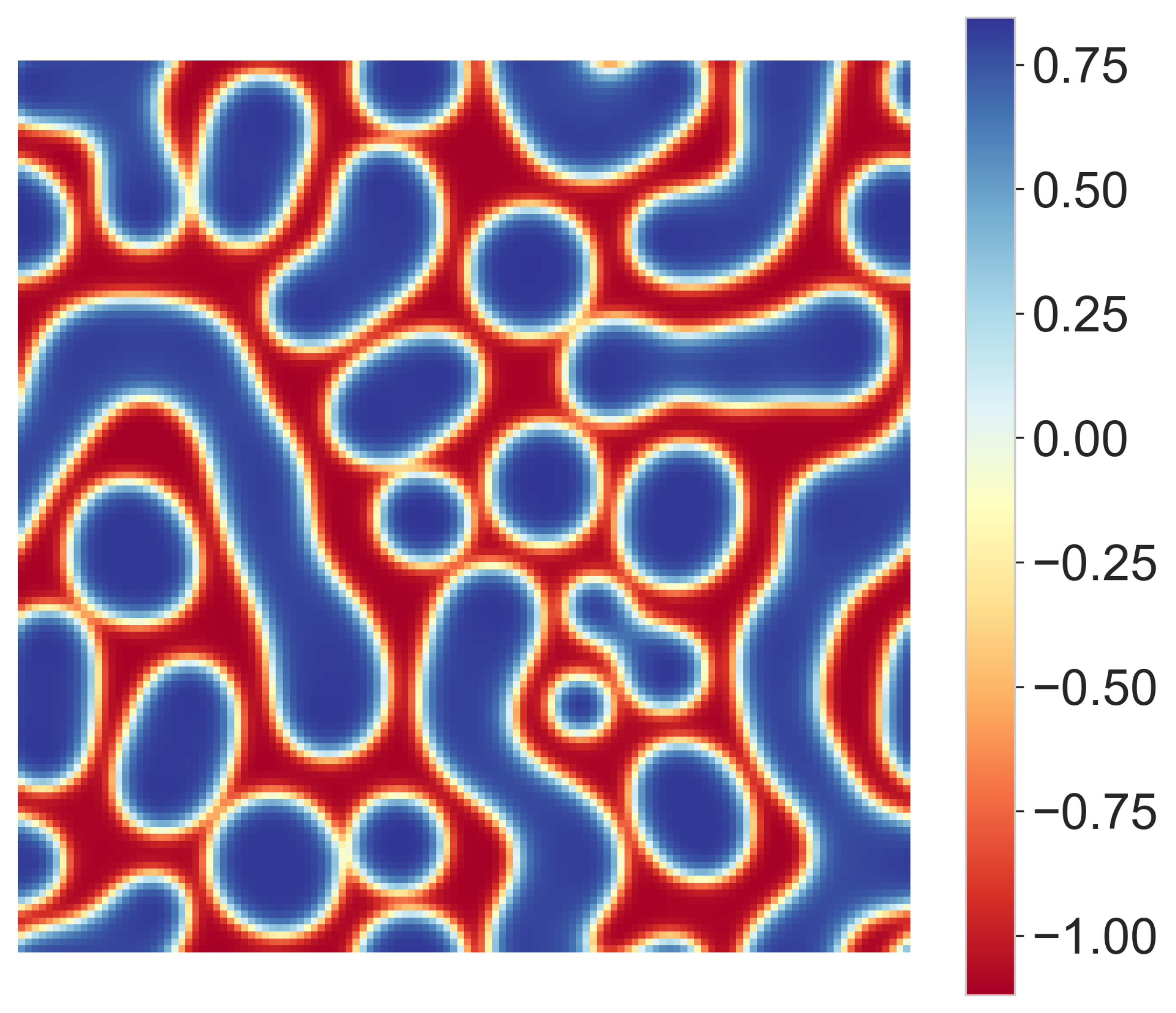}
  \caption{}
  \label{fig:6_act}
\end{subfigure}
	\caption{Density evolution snapshots of the Active model B .  (a) at t=0, (b) at t=100, (c) at t=200, (d) at t=300, (e) at t=400, (f) at t=500. The color bar represents the value of local density $\phi$.}
\label{fig:active}
\end{figure}

{\bf Active model B (AMB)} :- In this section, we will discuss the results of active model B. This model has a constant  activity ($\lambda=1.0$)
all over the system with box size $L=256$. 
%Simulations of Active Model B have been studied computationally, 
This study is performed to  understand the  comparison between the constant  activity parameter (AMB) and Gaussian distribution of activity parameter (IAMB). 
We observe an interesting difference in the kinetics of domain formation and steady state structure of domains. Fig.  \ref{fig:active} shows the real space 
snapshot of density at time steps = $0$, $100$, $200$, $300$, $400$, $500$. Starting from random homogeneous density, as time progress, A-particle regions start to phase separate and
phase separation happens with the formation of isolated domains. \cite{ncomms5351}.  

\subsection{ Time evolution of density for inhomogeneous activity (IAMB)} In this section, we will discuss the results of  
Gaussian activity of intensity, {\em i.e.},  $G=[1.0,L/4]$, $G=[2.0,L/4]$, $G=[3.0,L/4]$. In this case at the middle
of the box, the strength of the activity is highest and as we go away it spreads like a normal distribution.
%of I employed multi-variate normal distribution for varying the activity with peak value at the center of the system.
When we include the distribution of  activity, we observe that the A-particle accumulates at the center, where there is higher activity.
Accumulation of particles decreases as we move away from the center of the system. When we enter into zero  activity space, we can observe that particles form connected domains  from walls 
of system to the boundary of activity distribution. This connected domain structure is result of the passive phase separation mechanism or passive model B \cite{chc2,chc,chc3}. 
The density of both particles are conserved in this system. There are some droplets formation at the boundary of the distribution of constant activity, which is characteristic of active model B.
Fig. \ref{fig:active}.
\vspace{5cm}

\begin{figure}[H]
    \centering % <-- added
\begin{subfigure}{0.25\textwidth}
  \includegraphics[width=\linewidth]{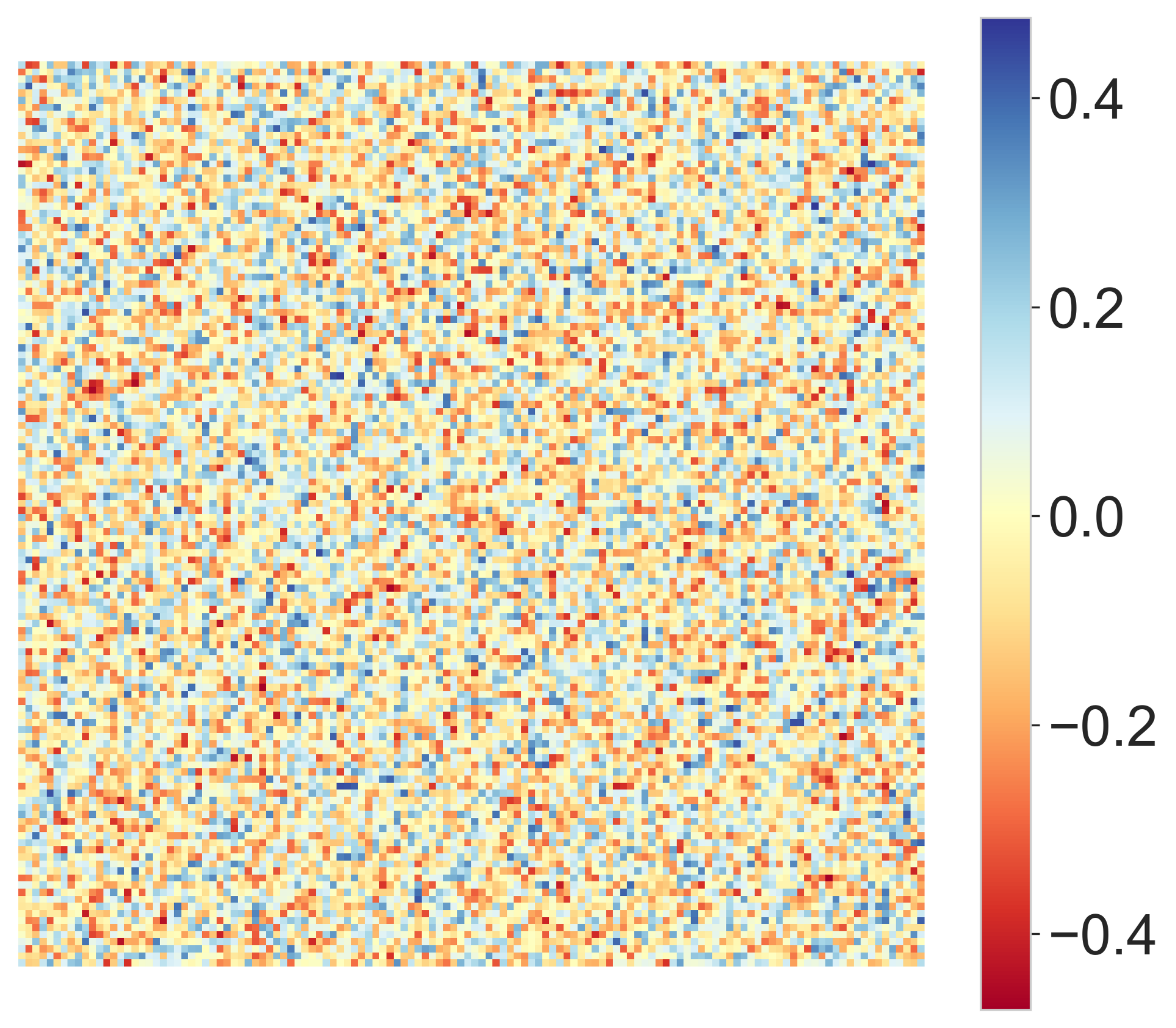}
  \caption{}
  \label{fig:1_1}
\end{subfigure}\hfil % <-- added
\begin{subfigure}{0.25\textwidth}
  \includegraphics[width=\linewidth]{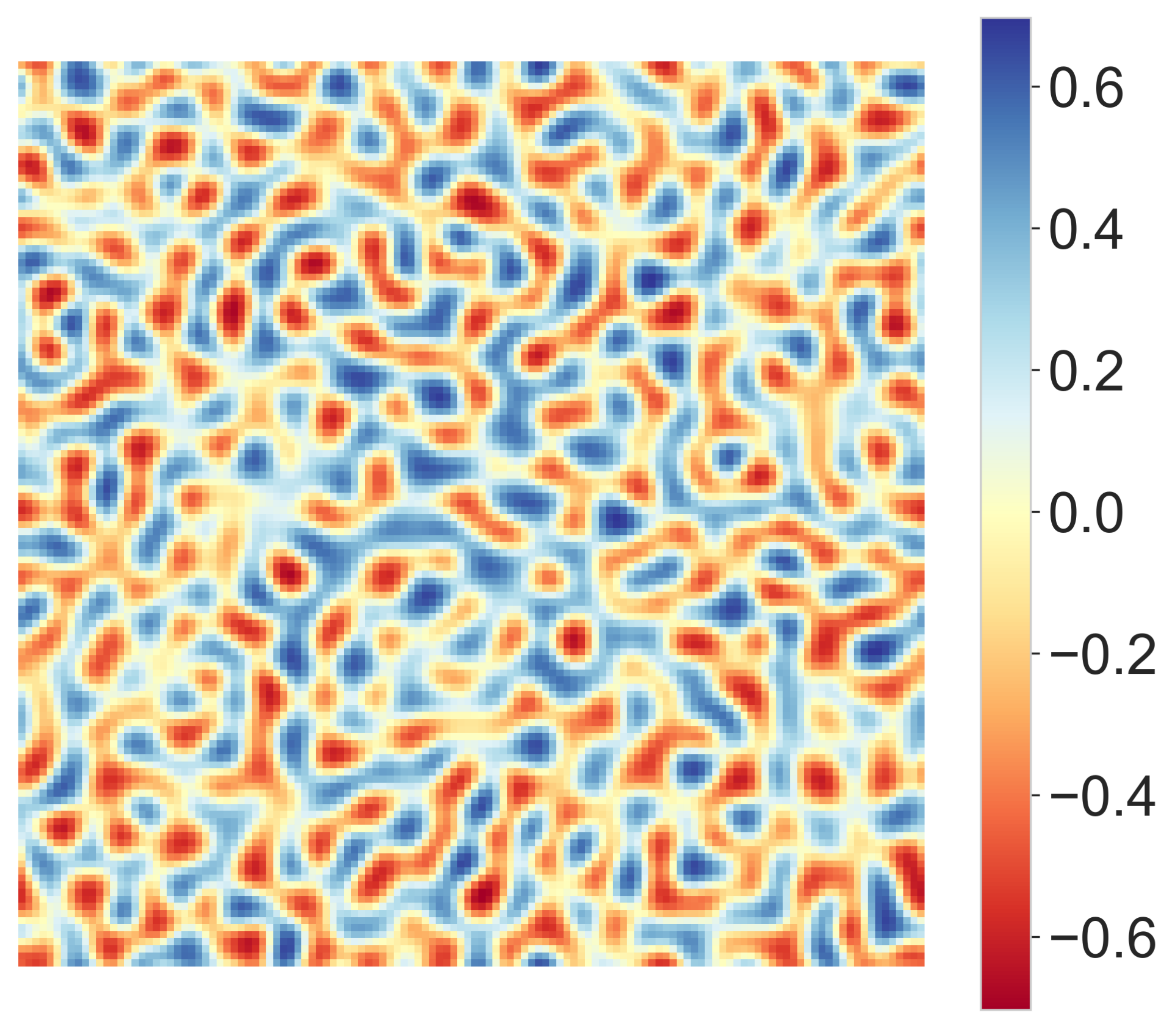}
  \caption{}
  \label{fig:2_1}
\end{subfigure}\hfil % <-- added
\begin{subfigure}{0.25\textwidth}
  \includegraphics[width=\linewidth]{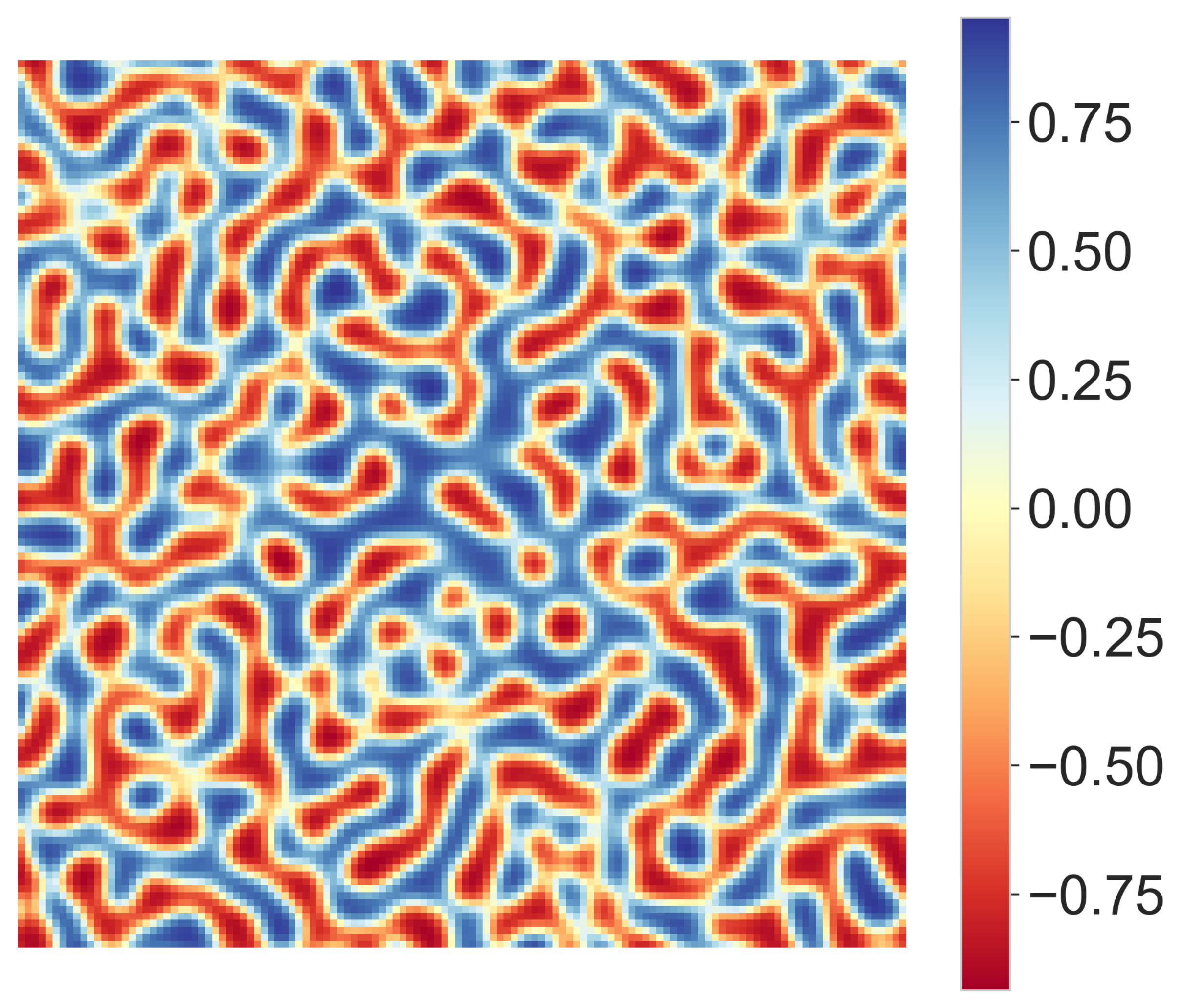}
  \caption{}
  \label{fig:3_1}
\end{subfigure}

\medskip
\begin{subfigure}{0.25\textwidth}
  \includegraphics[width=\linewidth]{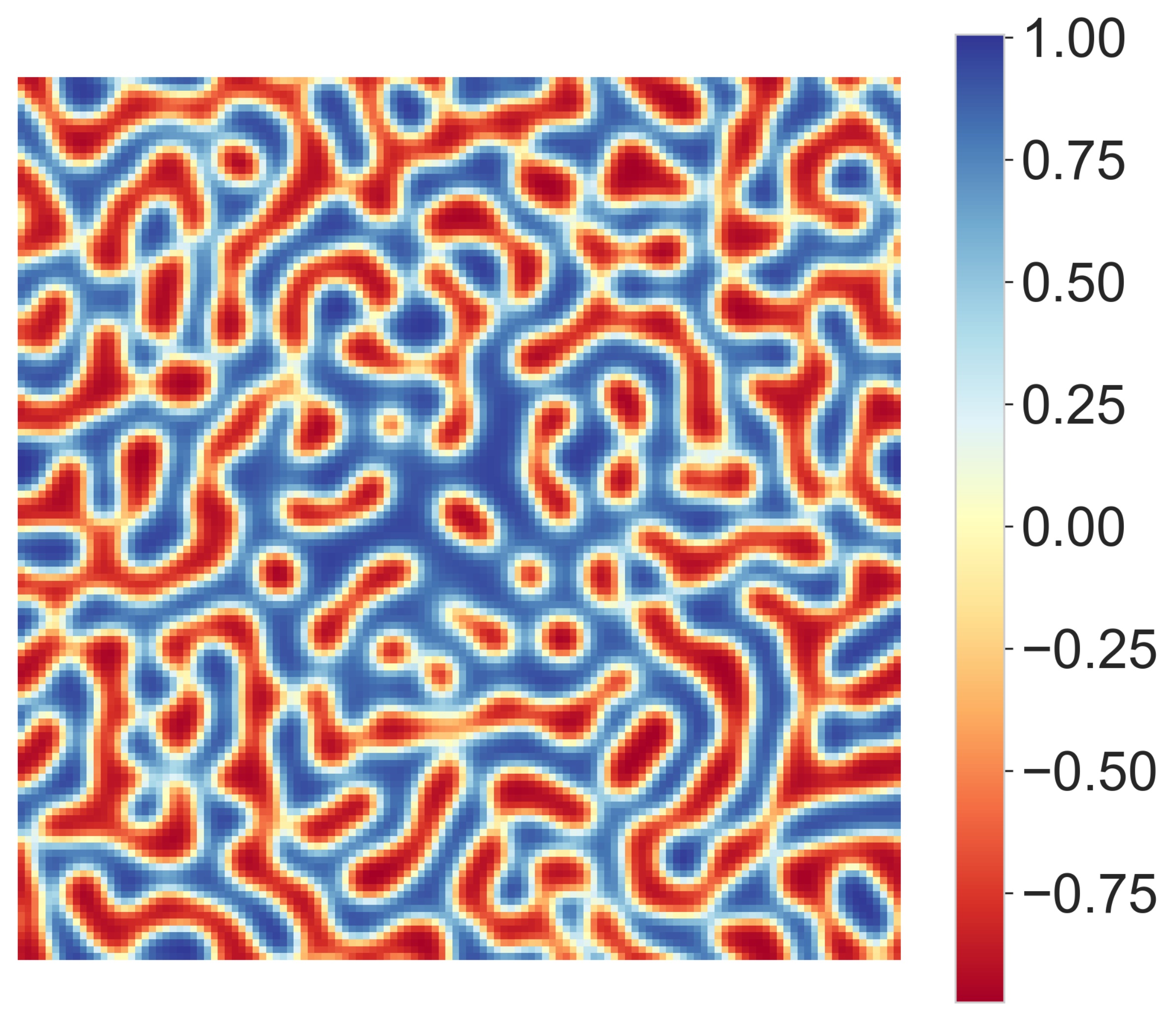}
  \caption{}
  \label{fig:4_1}
\end{subfigure}\hfil % <-- added
\begin{subfigure}{0.25\textwidth}
  \includegraphics[width=\linewidth]{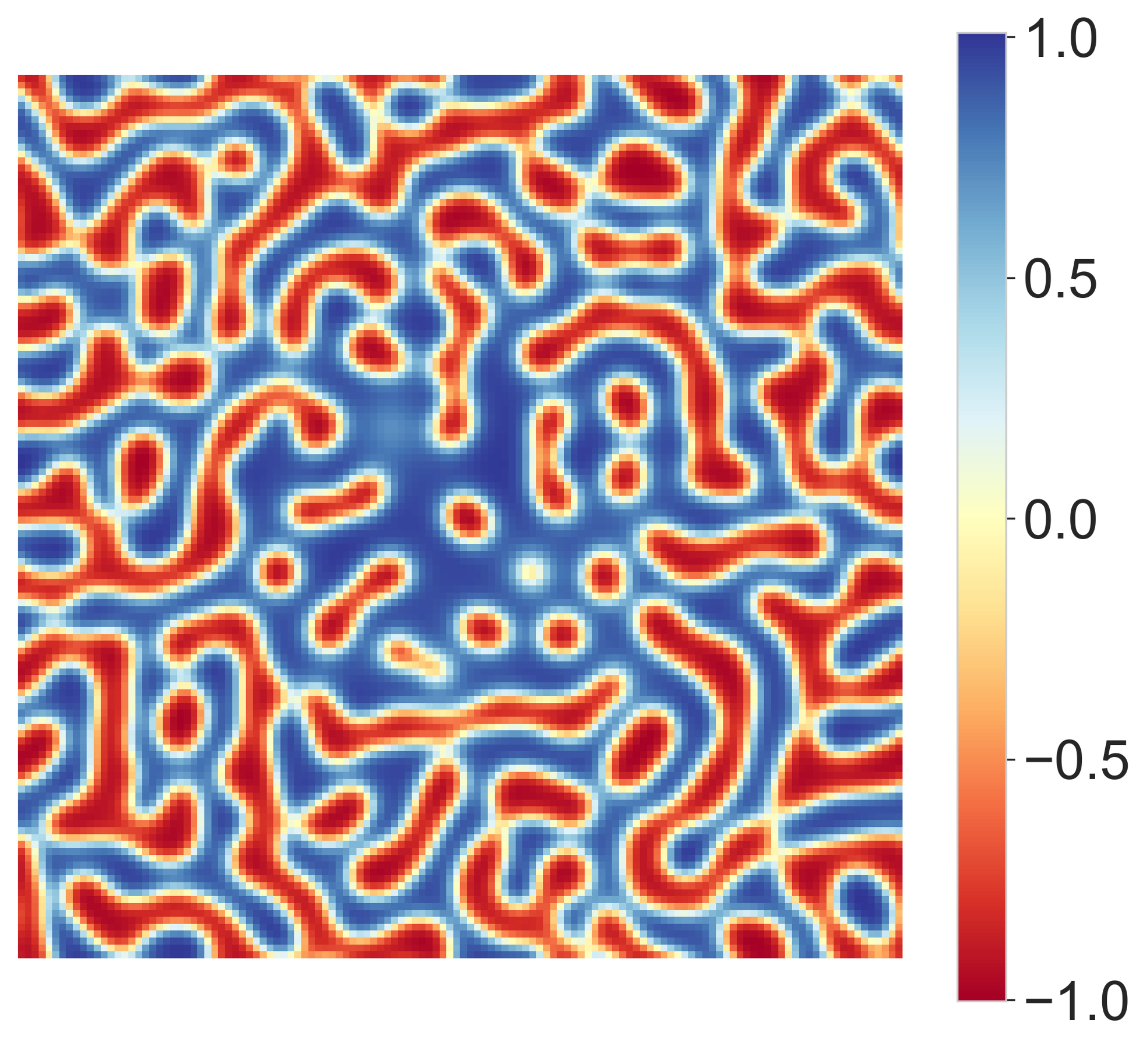}
  \caption{}
  \label{fig:5_1}
\end{subfigure}\hfil % <-- added
\begin{subfigure}{0.25\textwidth}
  \includegraphics[width=\linewidth]{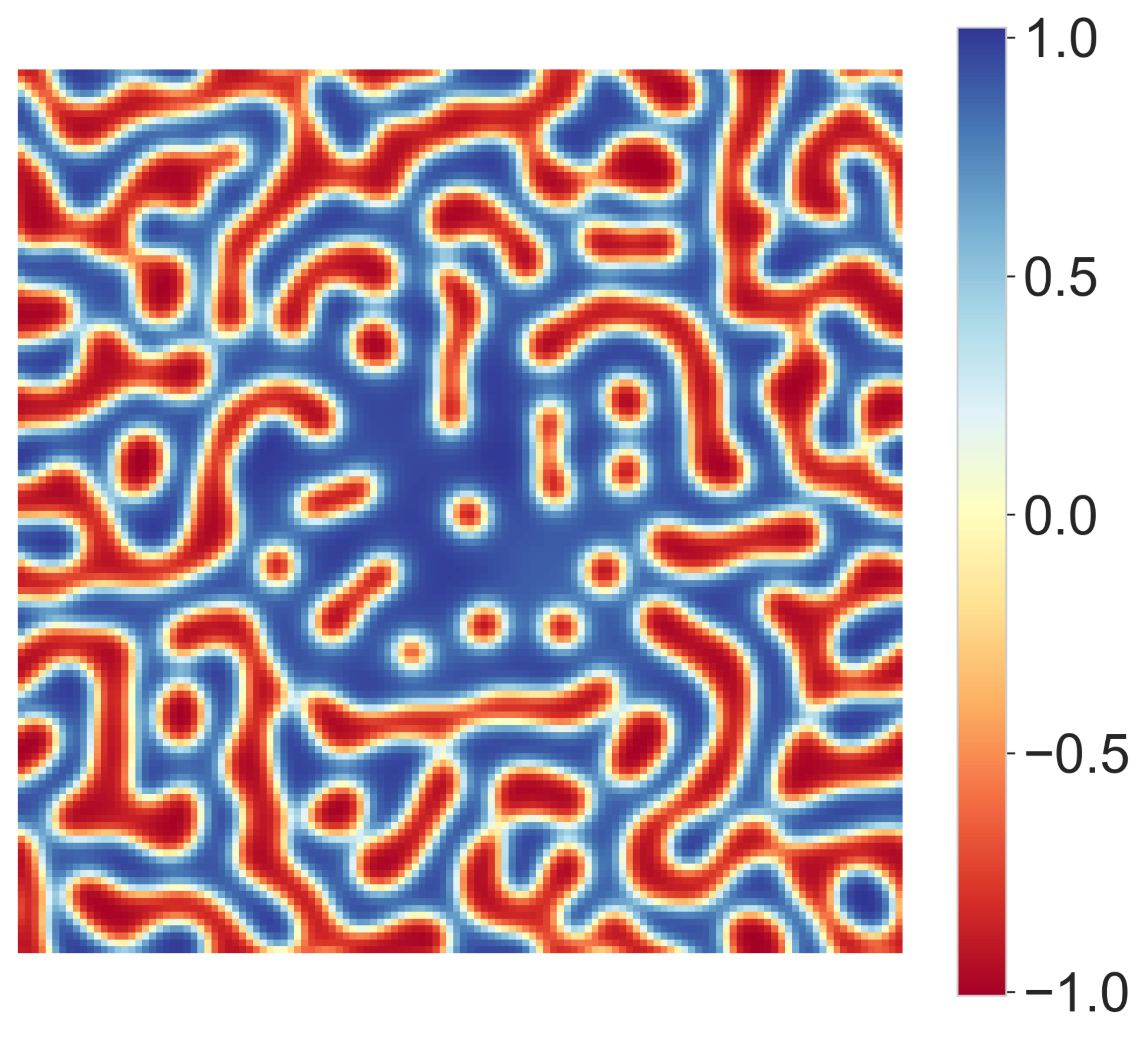}
  \caption{}
  \label{fig:6_1}
\end{subfigure}
	\caption{Evolution snapshots of the active model B with Gaussian distribution of activity with $\lambda_0=1.0$. (a) at t=0, (b) at t=15, (c) at t=25, (d) at t=35, (e) at t=40, (f) at t=50. The color bar have the same meaning as in Fig. \ref{fig:active}}
\label{fig:images_1}
\end{figure}

{\bf Case 1: Gaussian Distribution of activity with $\lambda_0=1$ :} In this case we consider the lowest intensity distribution i.e., ($\lambda_0=1$) . In Fig.  \ref{fig:images_1}, we attached snapshots of density evolution. 
%Essential parameters are $dx=0.0$, $dt=0.01$, $\lambda=[0.0.16,32]$, $t=5000$.   
% Simulation

In Fig. \ref{fig:images_1}, we have snapshots of density evolution with time. When the intensity of distribution is $\lambda_0=1.0$, we observe small-scale accumulation of A-particles at the center of the box. In Fig. \ref{fig:1_1}, particles are randomly distributed in system, so we observe scattered density. With time, we can see the accumulation of A-density particles at the center of system. Droplets of B-particles are also formed besides the accumulation of  A-particles in activity region. The growth of the accumulation is shown in the Fig. \ref{256_len}.
\vspace{5cm}

\begin{figure}[H]
    \centering % <-- added
\begin{subfigure}{0.25\textwidth}
  \includegraphics[width=\linewidth]{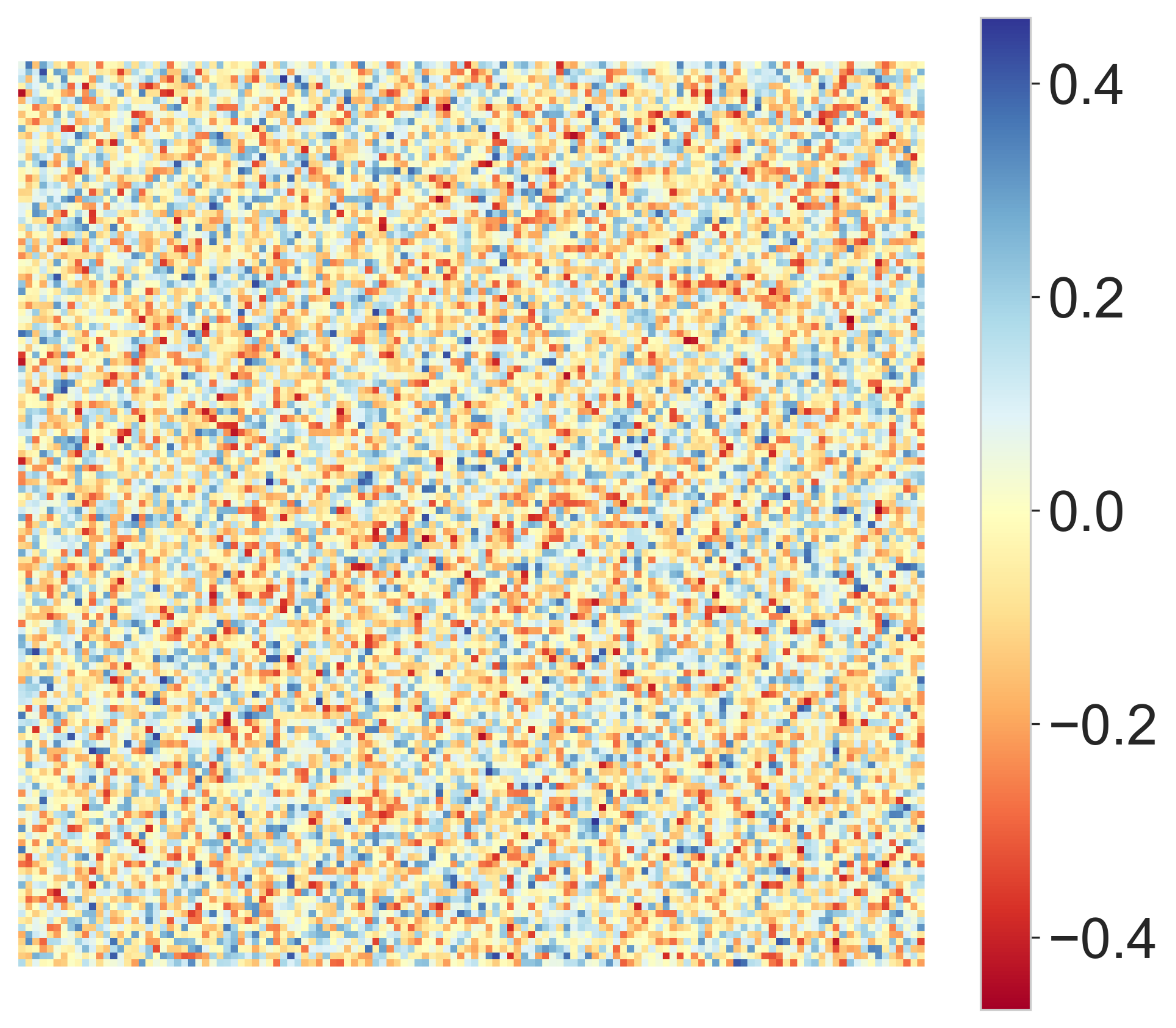}
  \caption{}
  \label{fig:1_2}
\end{subfigure}\hfil % <-- added
\begin{subfigure}{0.25\textwidth}
  \includegraphics[width=\linewidth]{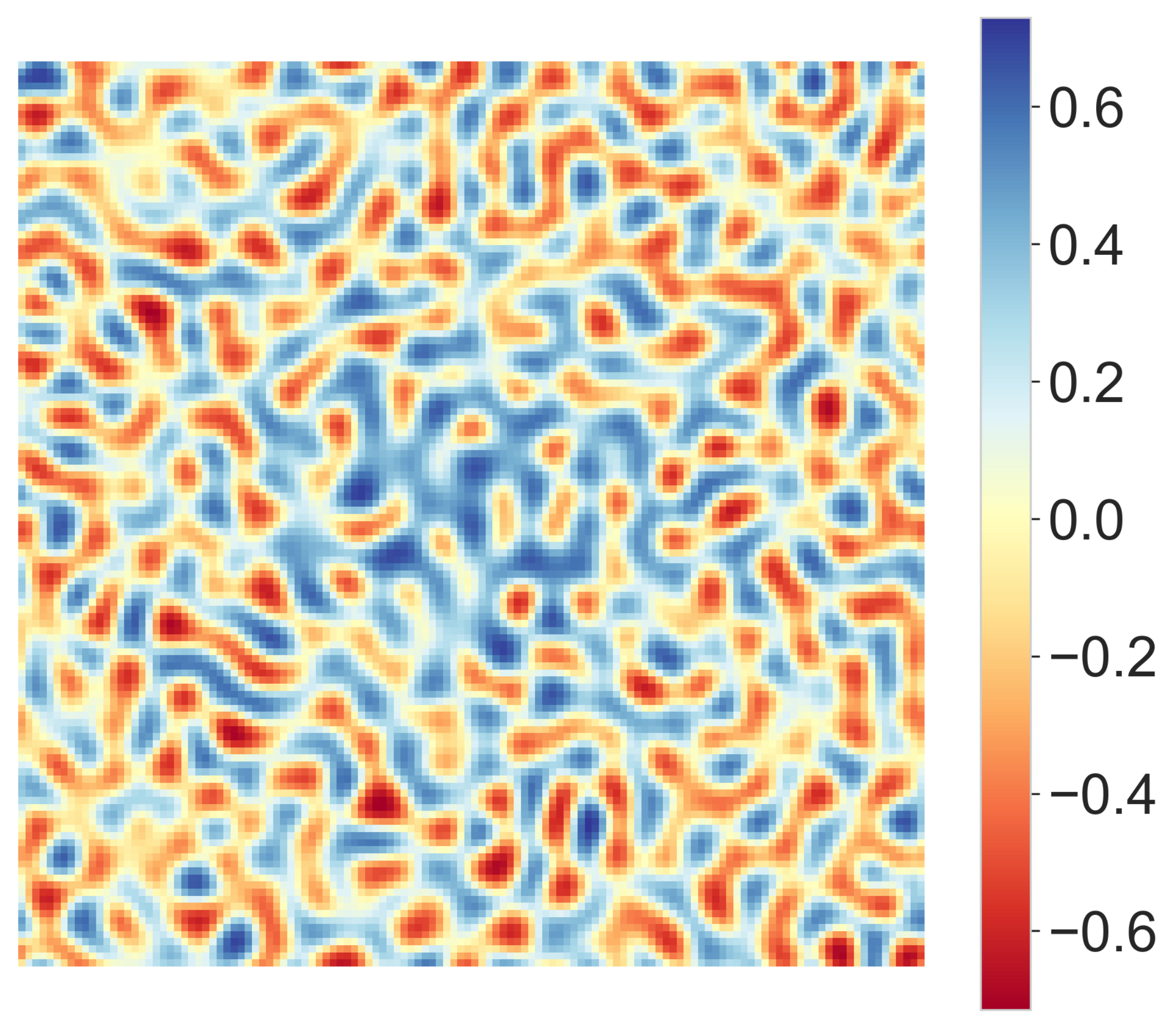}
  \caption{}
  \label{fig:2_2}
\end{subfigure}\hfil % <-- added
\begin{subfigure}{0.25\textwidth}
  \includegraphics[width=\linewidth]{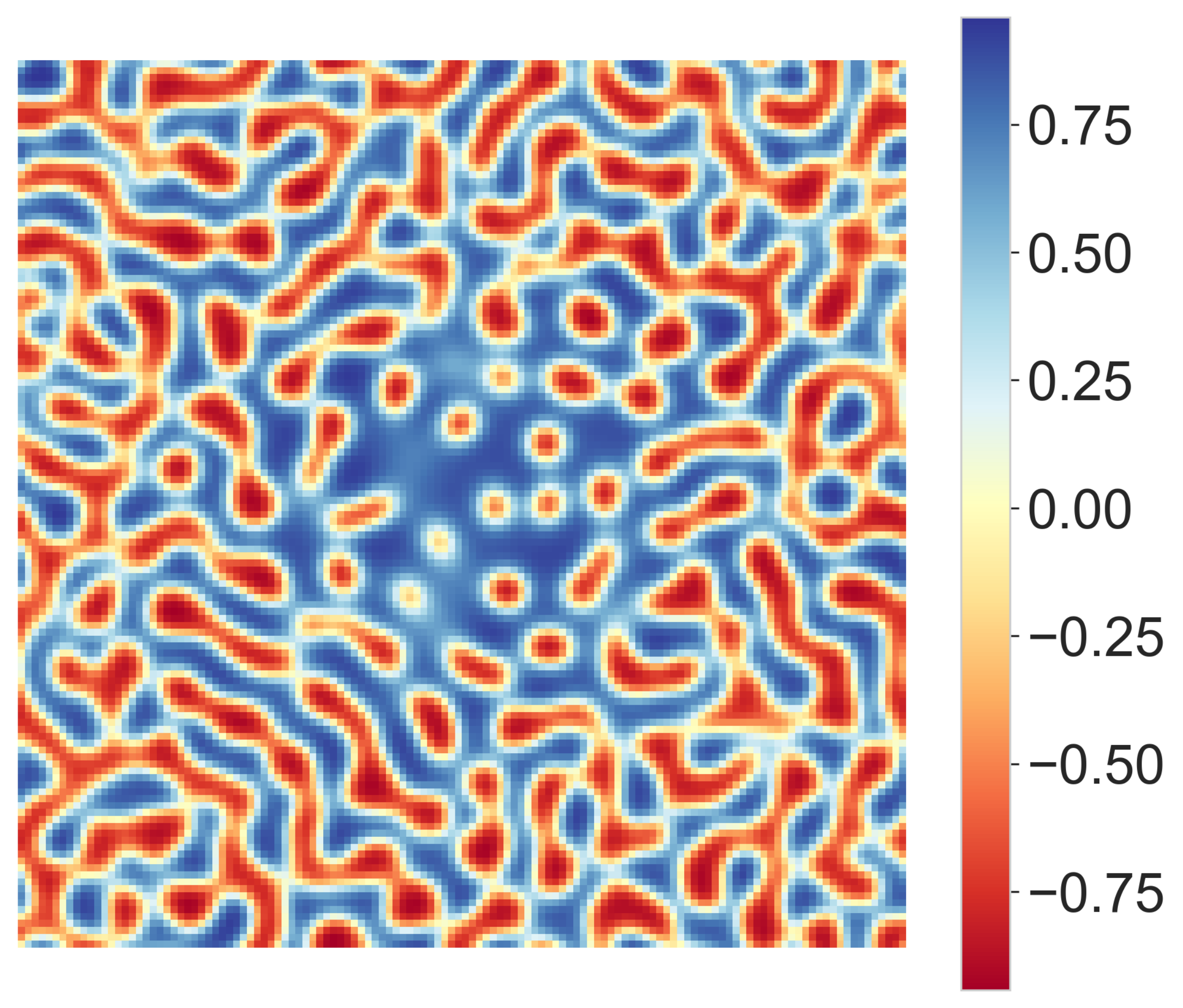}
  \caption{}
  \label{fig:3_2}
\end{subfigure}

\medskip
\begin{subfigure}{0.25\textwidth}
  \includegraphics[width=\linewidth]{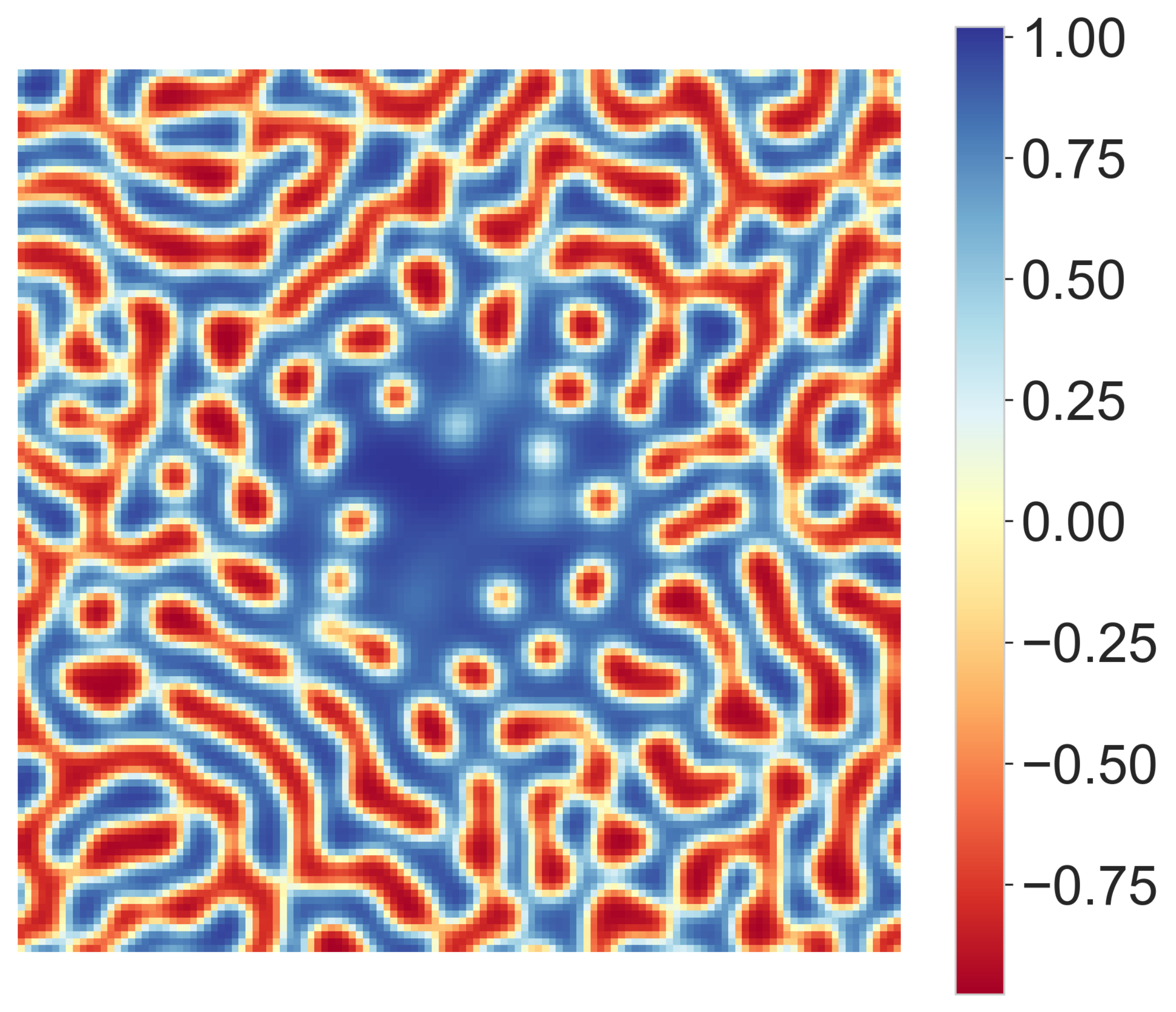}
  \caption{}
  \label{fig:4_2}
\end{subfigure}\hfil % <-- added
\begin{subfigure}{0.25\textwidth}
  \includegraphics[width=\linewidth]{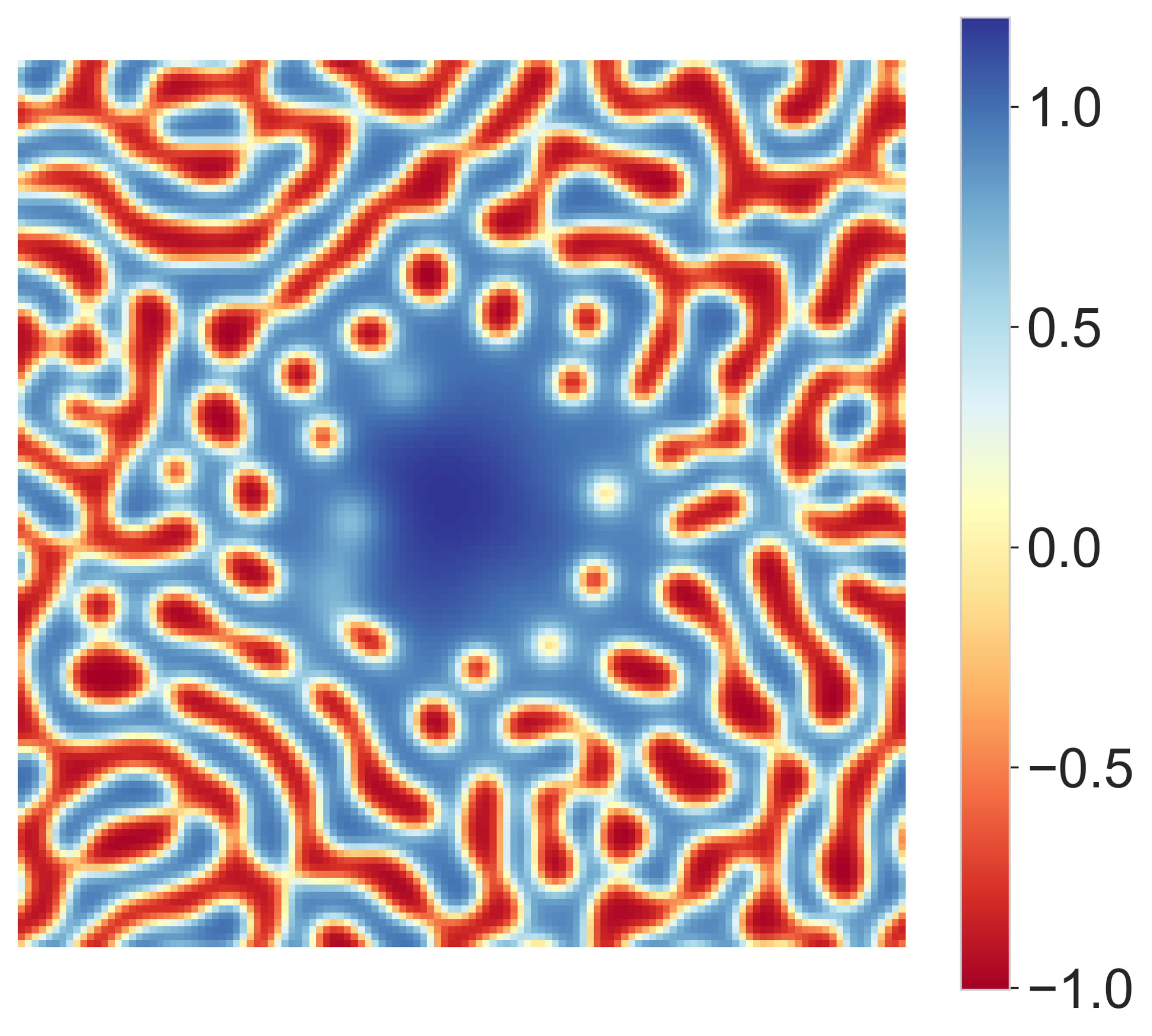}
  \caption{}
  \label{fig:5_2}
\end{subfigure}\hfil % <-- added
\begin{subfigure}{0.25\textwidth}
  \includegraphics[width=\linewidth]{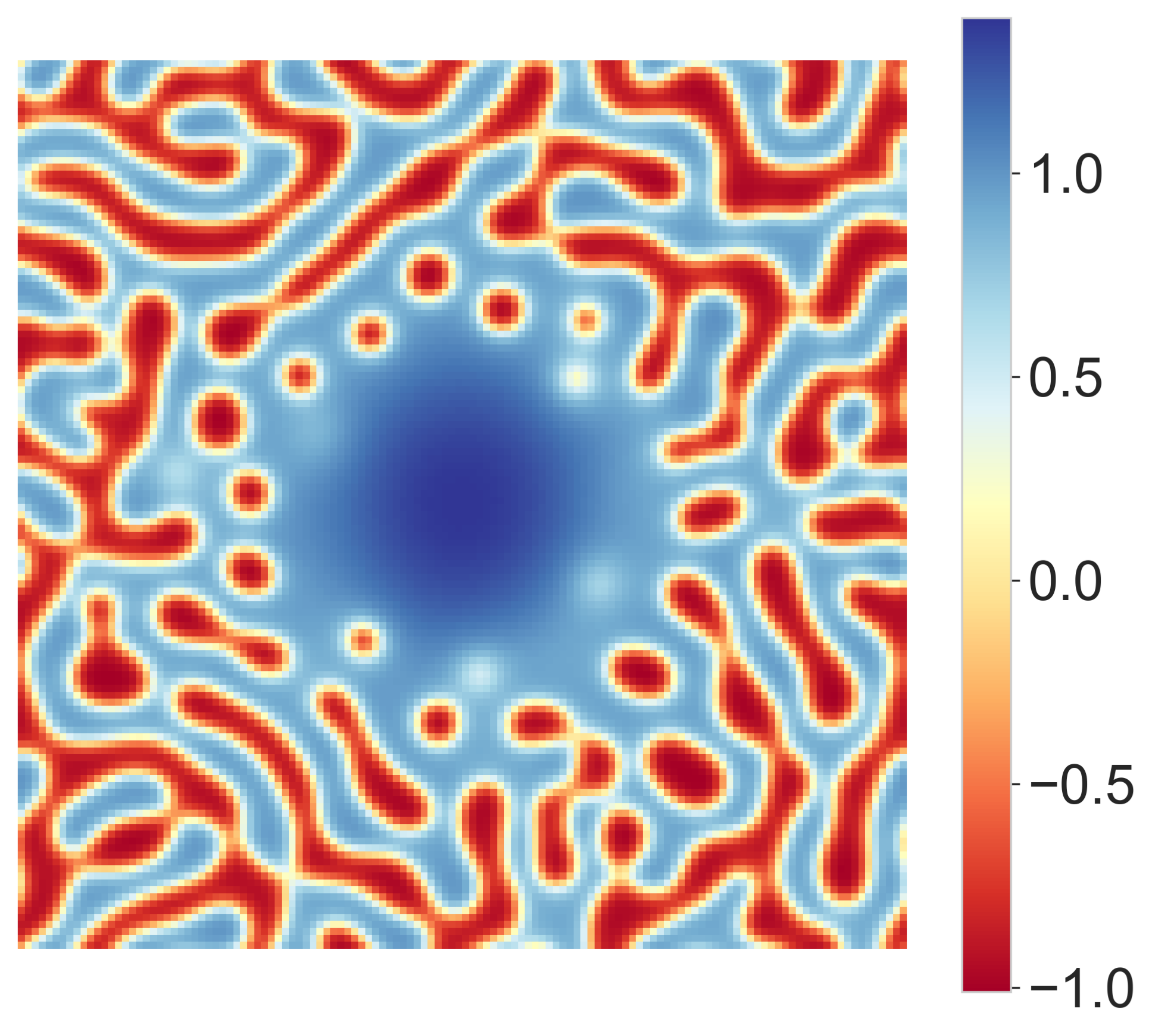}
  \caption{}
  \label{fig:6_2}
\end{subfigure}
	\caption{Evolution snapshots of the active model B with Gaussian distribution of activity with $\lambda_0=2.0$. (a) at t=0, (b) at t=15, (c) at t=25, (d) at t=35, (e) at t=40, (f) at t=50. The color bar have the same meaning as in Fig. \ref{fig:active}}
\label{fig:images_2}
\end{figure}

{\bf Case 2: Gaussian Distribution of activity with  $\lambda_0=2$ :} In this case we consider the  intensity distribution. In Fig. \ref{fig:images_2} we show snapshots of density evolution of the system. 
%The parameters are $dt=0.1$, $dx = 0.0$, $t=5000$, $\lambda=[0.032,32]$.

%Simulation
In Fig. \ref{fig:images_2} we can observe the evolution of the system. 
When the intensity of distribution is $\lambda_0=2.0$, we observe more  accumulation of A-particles at the center where there is maximum value for activity.
Traversing from the center of system to the walls of system, there is decrease in particle accumulation and we observe B-particles forms connected 
domains outside the activity domain. Droplets of B-particles are also formed besides accumulation of A-particles. The growth of the accumulation is shown in Fig. \ref{256_len}.
\vspace{5cm}

\begin{figure}[H]
    \centering % <-- added
\begin{subfigure}{0.25\textwidth}
  \includegraphics[width=\linewidth]{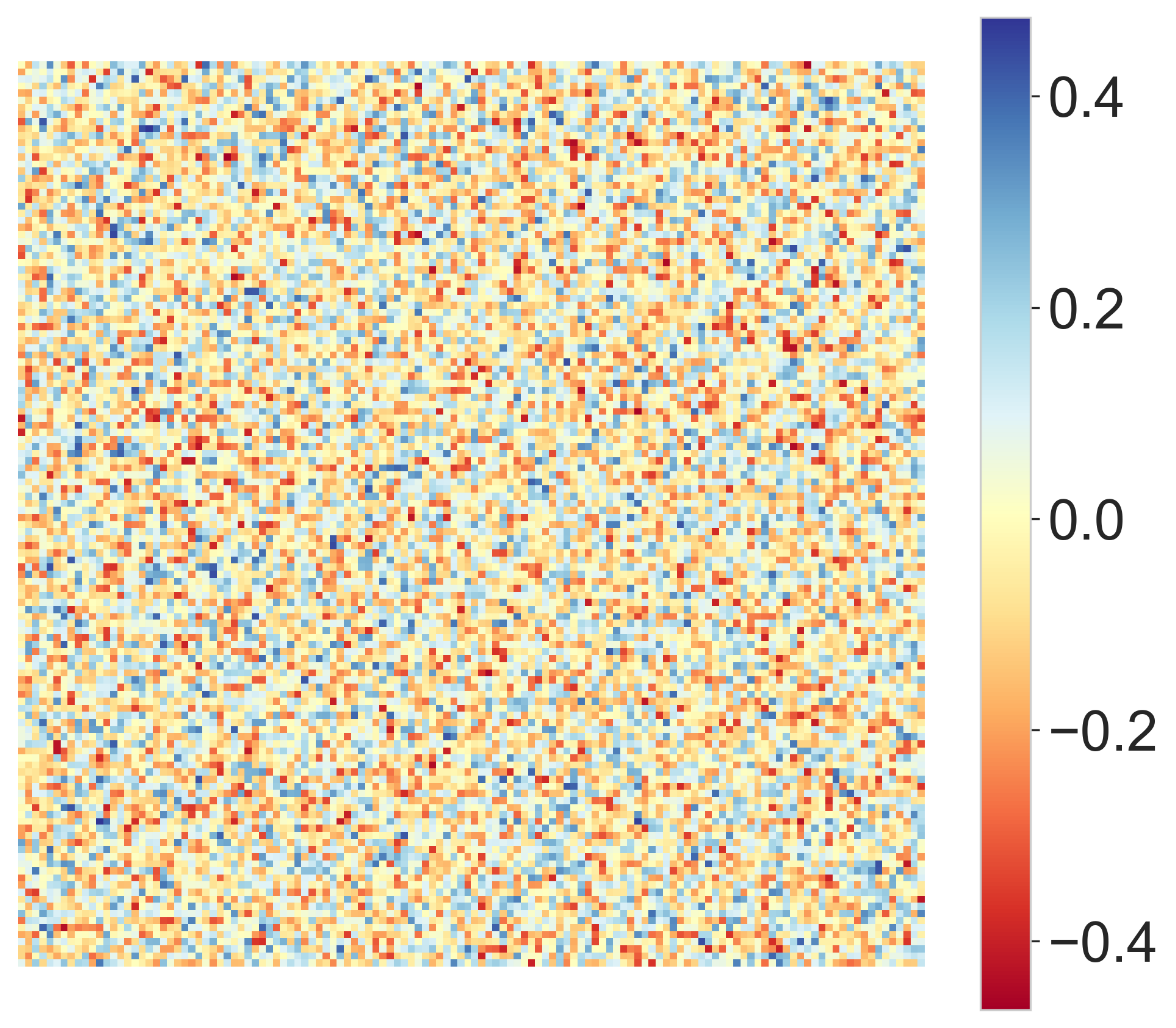}
  \caption{}
  \label{fig:1_3}
\end{subfigure}\hfil % <-- added
\begin{subfigure}{0.25\textwidth}
  \includegraphics[width=\linewidth]{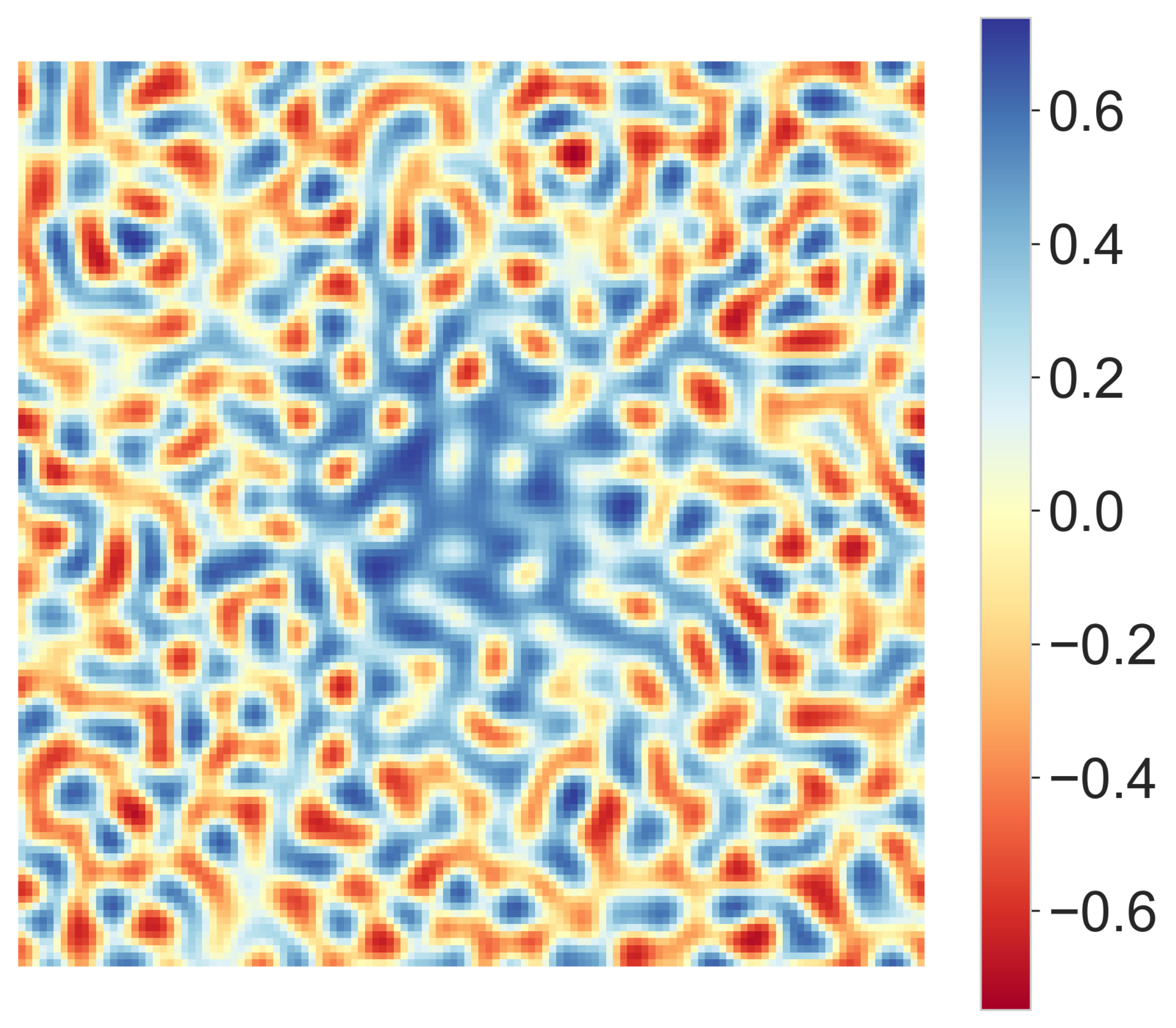}
  \caption{}
  \label{fig:2_3}
\end{subfigure}\hfil % <-- added
\begin{subfigure}{0.25\textwidth}
  \includegraphics[width=\linewidth]{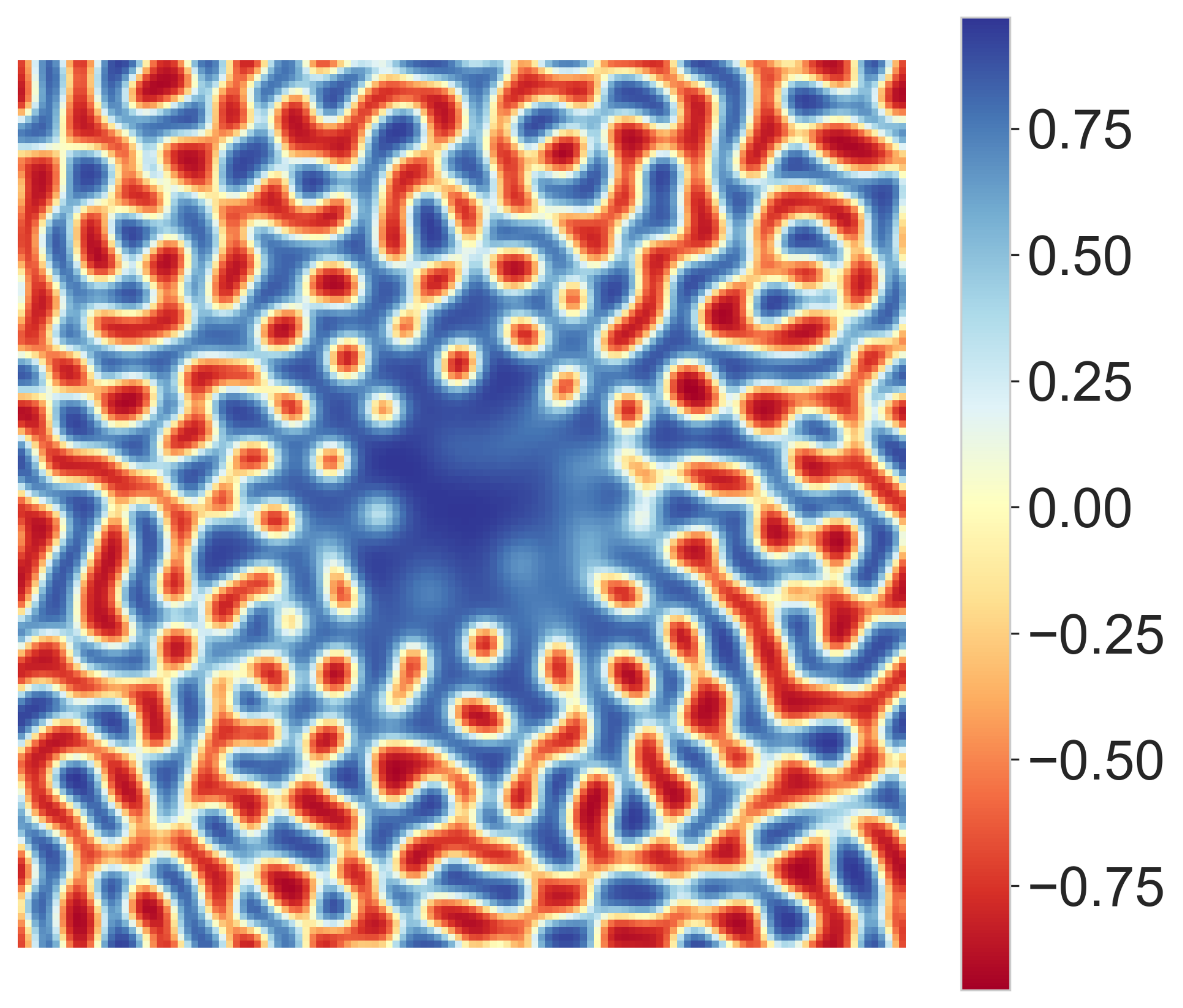}
  \caption{}
  \label{fig:3_3}
\end{subfigure}

\medskip
\begin{subfigure}{0.25\textwidth}
  \includegraphics[width=\linewidth]{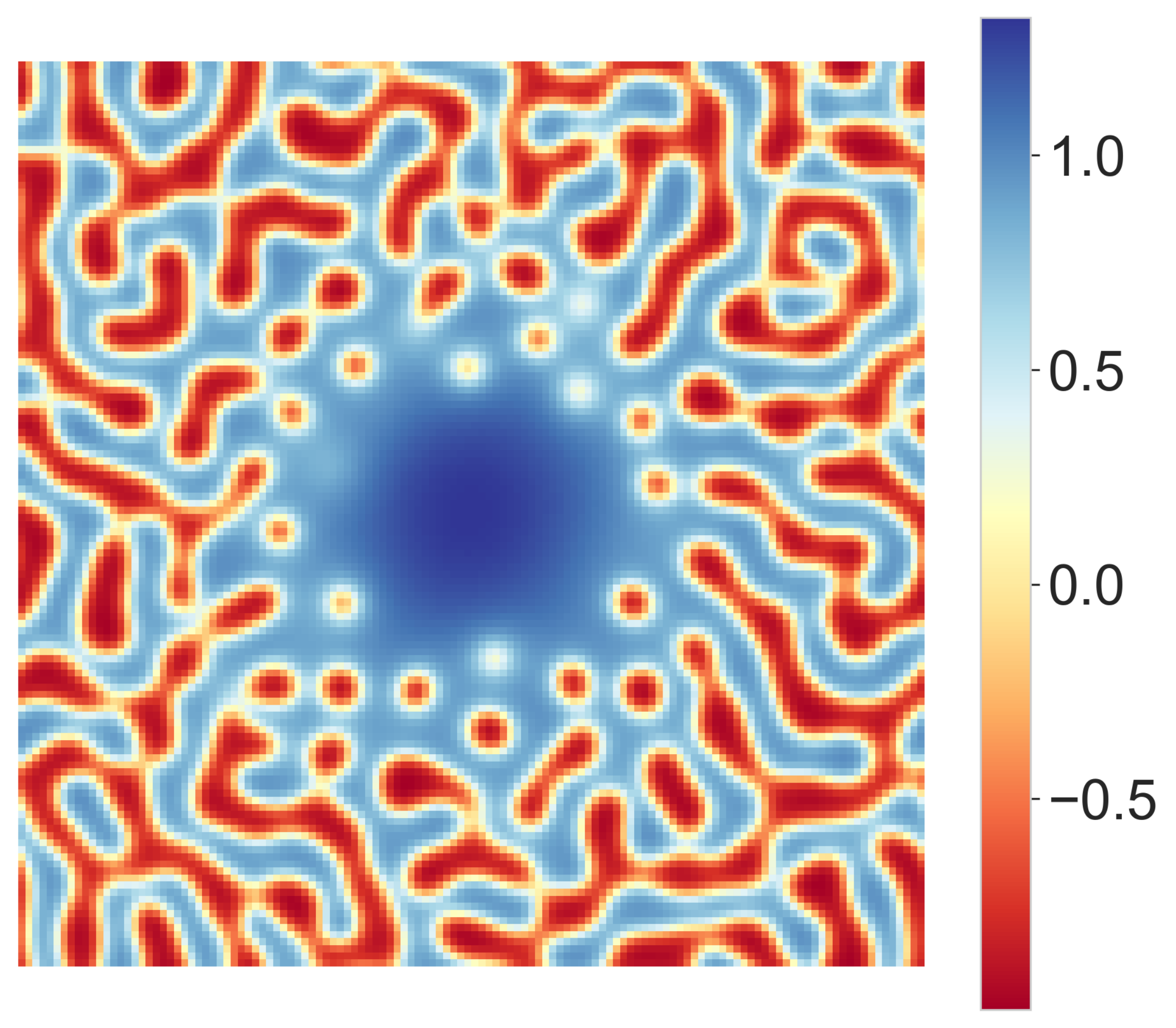}
  \caption{}
  \label{fig:4_3}
\end{subfigure}\hfil % <-- added
\begin{subfigure}{0.25\textwidth}
  \includegraphics[width=\linewidth]{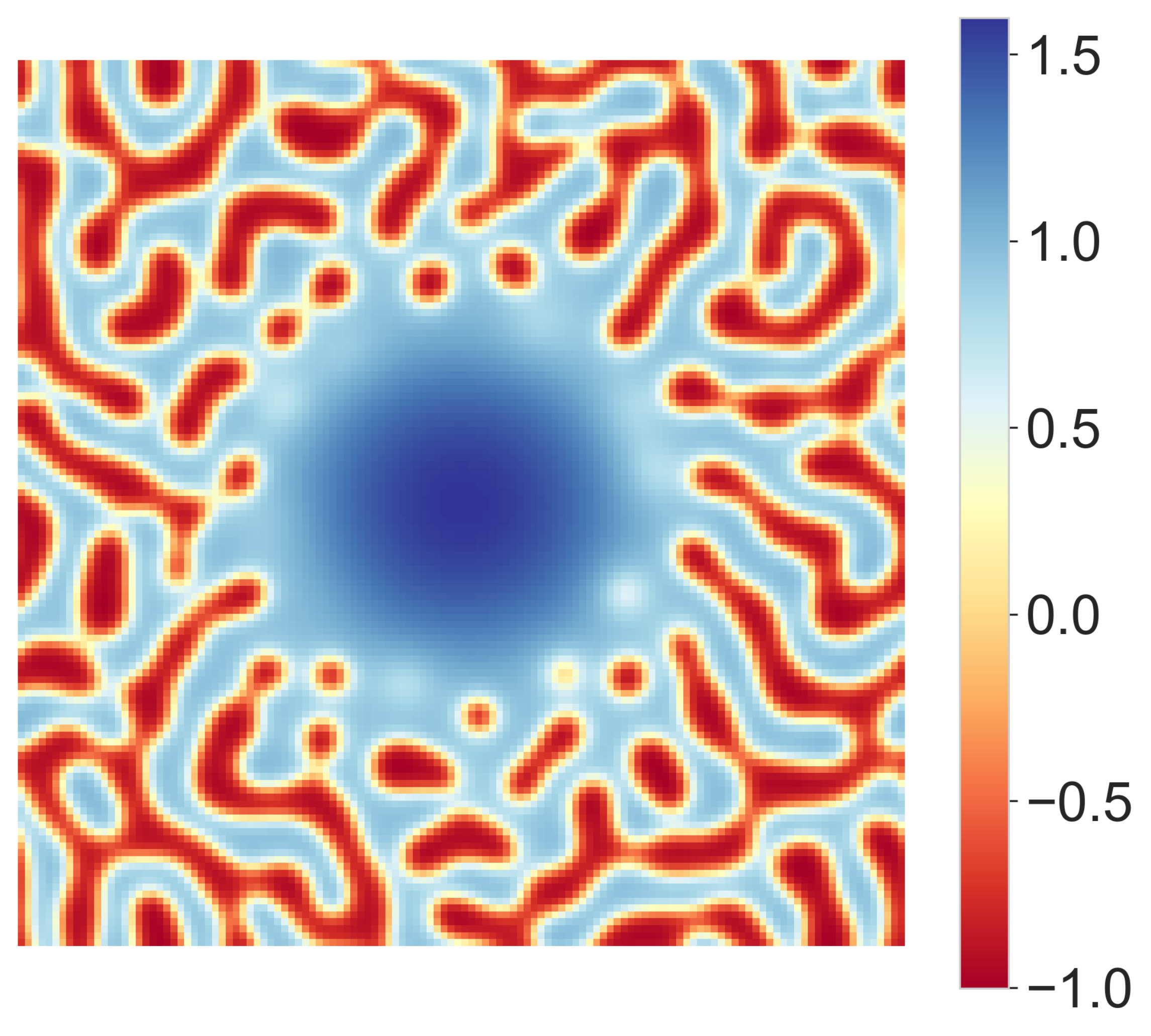}
  \caption{}
  \label{fig:5_3}
\end{subfigure}\hfil % <-- added
\begin{subfigure}{0.25\textwidth}
  \includegraphics[width=\linewidth]{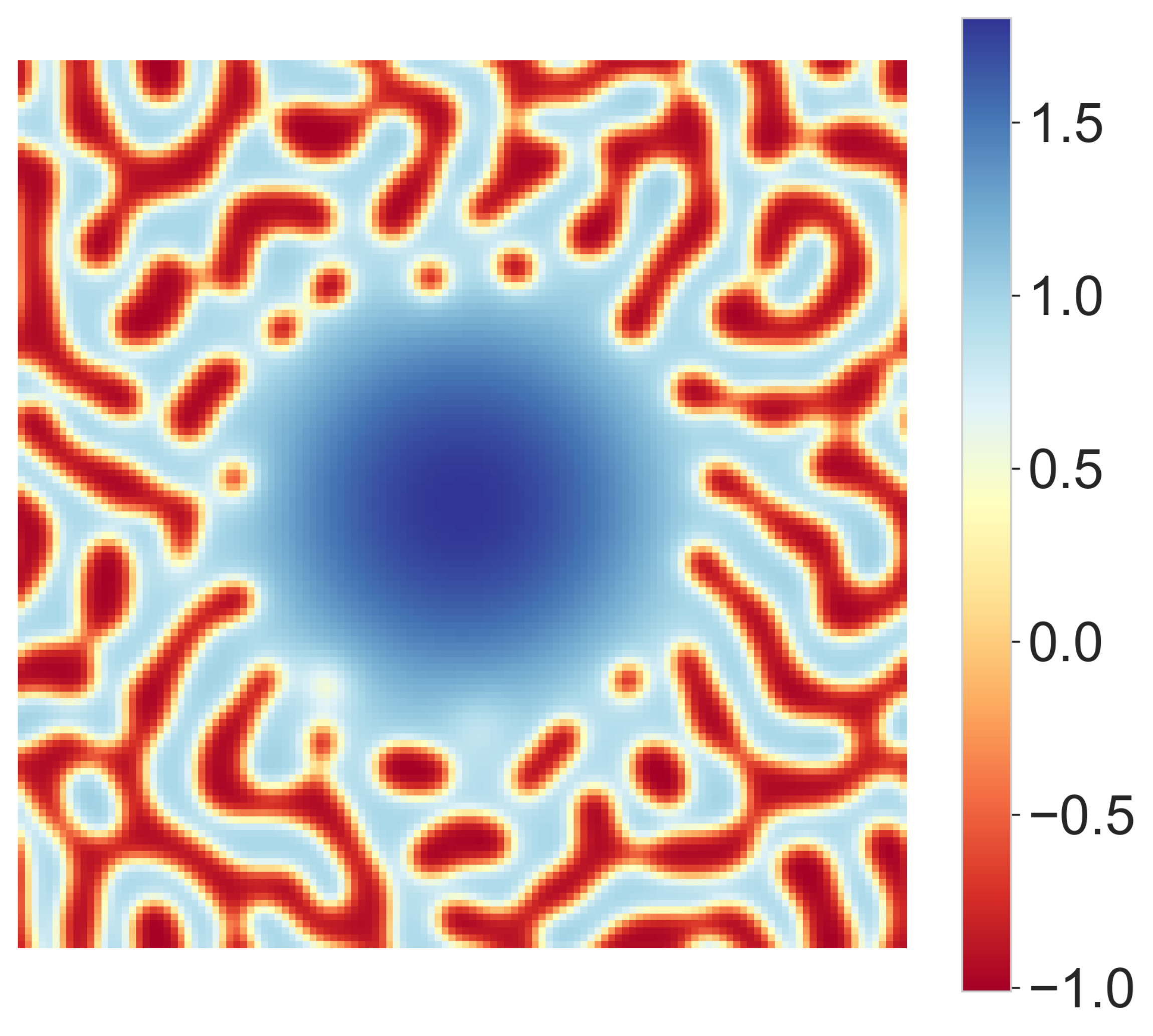}
  \caption{}
  \label{fig:6_3}
\end{subfigure}
	\caption{Evolution snapshots of the active model B with Gaussian distribution of activity with $\lambda_0=3.0$. (a) at t=0, (b) at t=15, (c) at t=25, (d) at t=35, (e) at t=40, (f) at t=50. The color bar have the same meaning as in Fig. \ref{fig:active}}
\label{fig:images_3}
\end{figure}

\begin{figure}[H]
    \centering % <-- added
\begin{subfigure}{0.5\textwidth}
  \includegraphics[width=\linewidth]{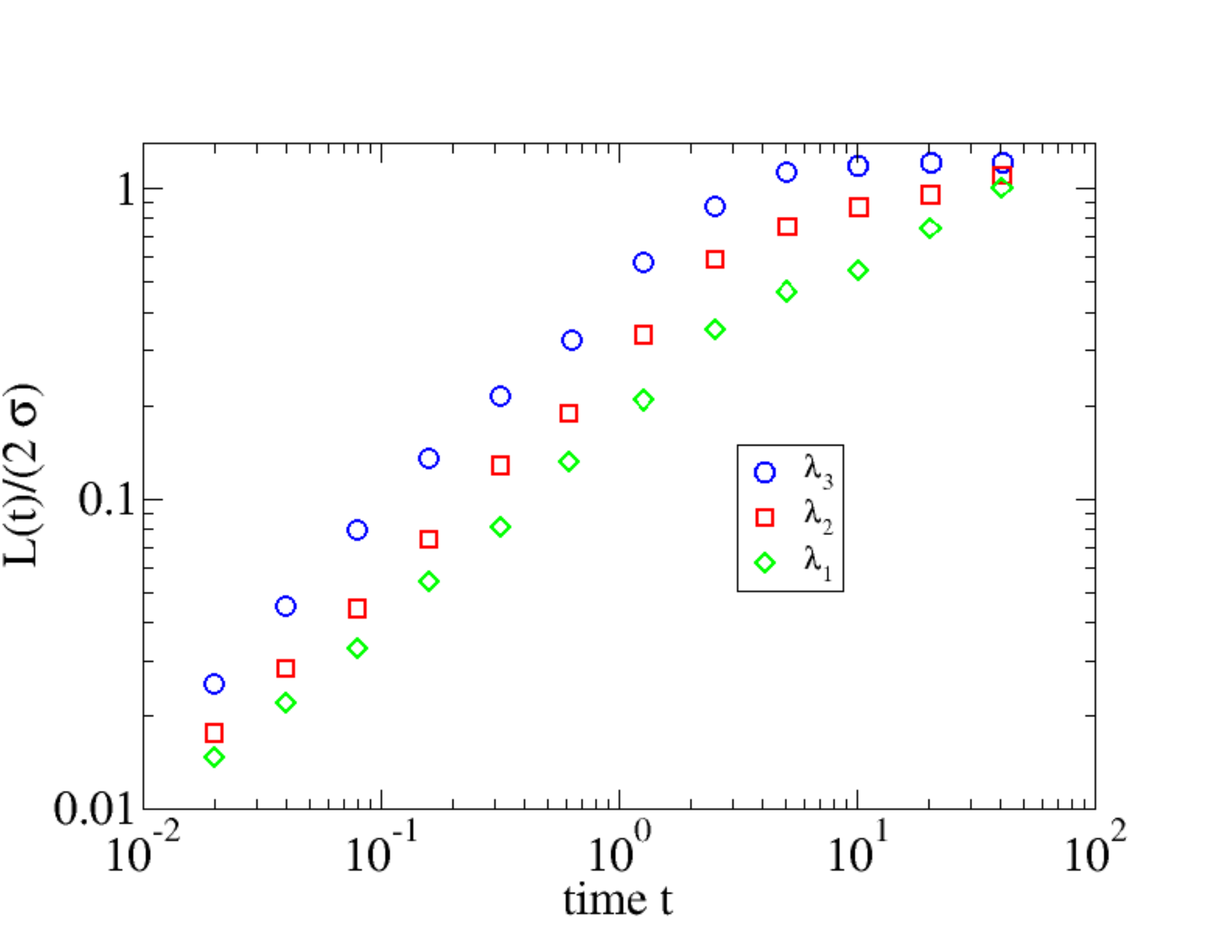}
  \caption{}
  \label{256_len}
\end{subfigure}\hfil % <-- added
\begin{subfigure}{0.5\textwidth}
  \includegraphics[width=\linewidth]{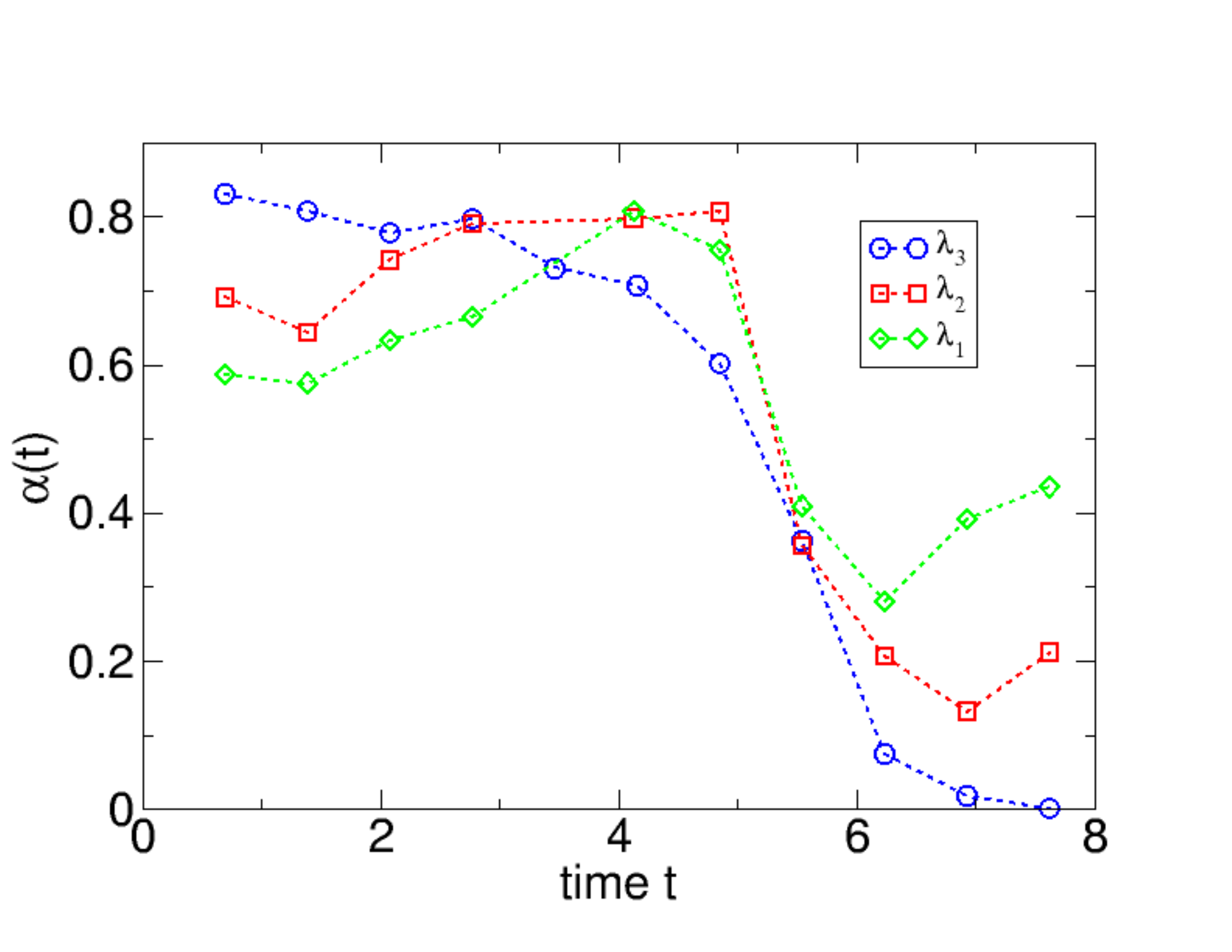}
  \caption{}
  \label{256_beta}
\end{subfigure}\hfil % <-- added
\begin{subfigure}{0.5\textwidth}
  \includegraphics[width=\linewidth]{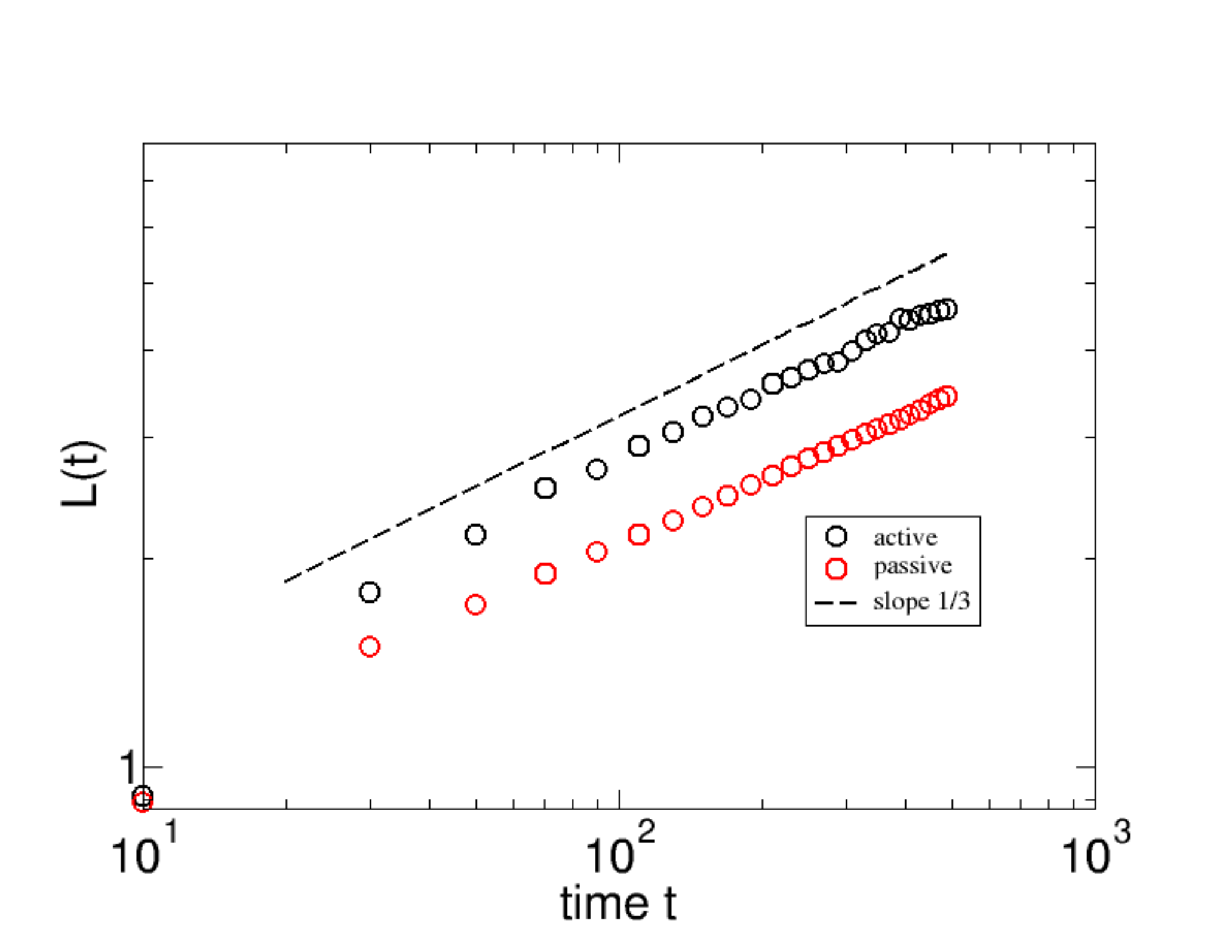}
  \caption{}
  \label{act_pas}
\end{subfigure}\hfil
\begin{subfigure}{0.5\textwidth}
  \includegraphics[width=\linewidth]{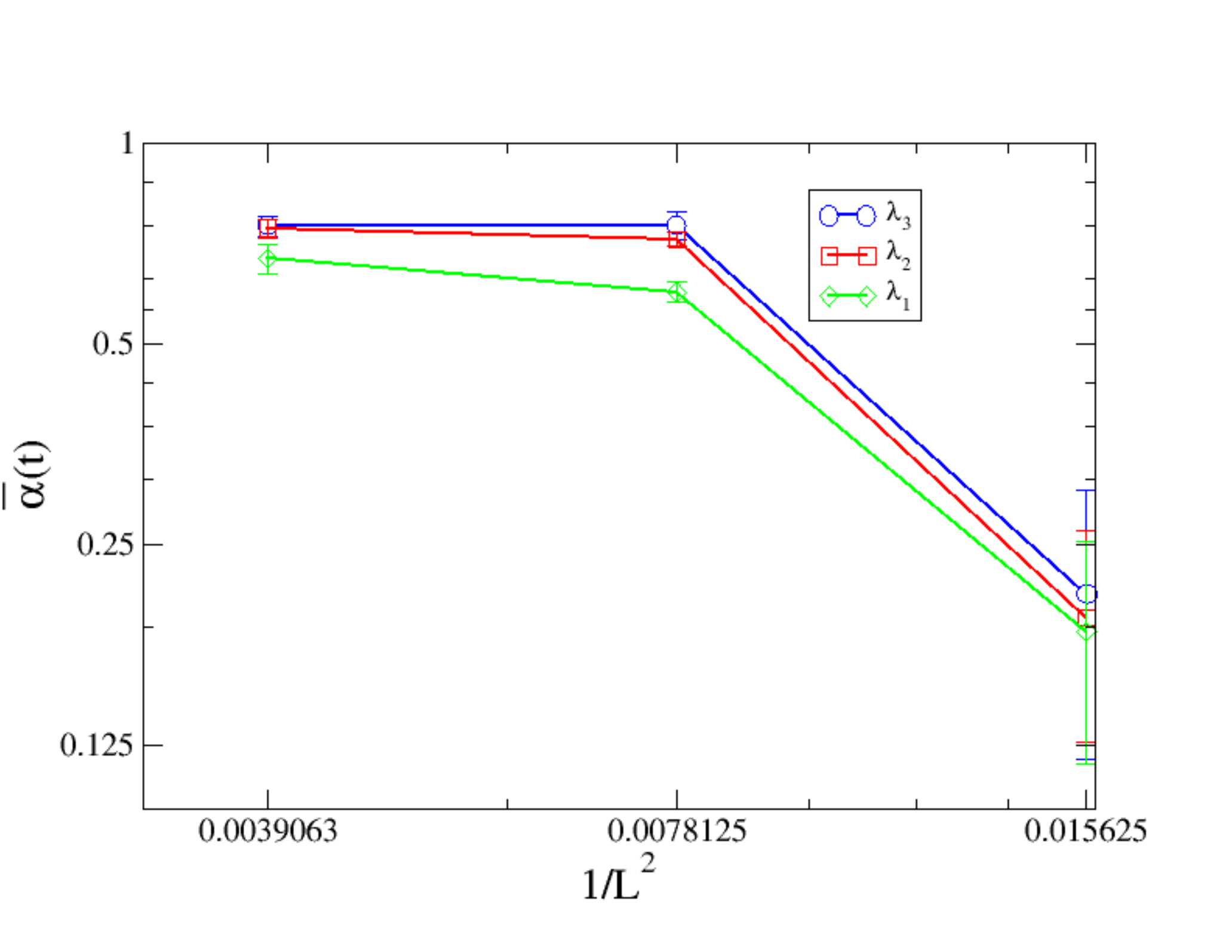}
  \caption{}
  \label{alpha}
\end{subfigure}
\medskip

	\caption{The Fig. \ref{256_len} shows the comparison plot of the growth of accumulation for 3 cases. As we see, the intensity 3 has faster growth followed by intensity 2 and then intensity 1. $\lambda_3$ represents the highest intensity and $\lambda_1$ represents the lowest intensity. The Fig. \ref{256_beta} shows the  $\alpha(t)$ plot for the length, which gives us the range of $\alpha$ for 3 cases. The Fig. \ref{act_pas} is the comparison length plot with time for active model and passive model.The Fig. \ref{alpha} shows the $\overline{\alpha}(t) vs \frac{1}{L^2}$ plot. We have taken average of the value of $\alpha(t)$ for each intensity i.e., ($\lambda_3$,$\lambda_2$,$\lambda_1$) for $64$, $128$, $256$ box sizes.} 
\label{fig:length}
\end{figure}

{\bf Case 3: Gaussian Distribution of activity with $\lambda_0=3 :$} In this case we consider the  intensity distribution. In Fig. \ref{fig:images_3} we show snapshots of density evolution of the system. 
%The parameters are $dt=0.1$, $dx = 0.0$, $t=5000$, $\lambda=[0.048,32]$.

%Simulation

 In Fig. \ref{fig:images_3} we can observe the evolution of the system. 
When the intensity of distribution is $\lambda_0=3.0$ which is the highest intensity we considered. Here in this system, we observe the accumulation of A-particles at 
the center where there is the maximum value for activity. %Very clearly at the center, there are no B-particles which convey us the accumulation of particle in high density region. 
Traversing from the center of the system to the walls of the system, there is a decrease in particle accumulation, and we observe B-particles form connected domains outside the activity region. 
Droplets of B-particles are also formed at the boundaries of the active region besides the accumulation of A-particles at the center of the system which is the phenomena of active model B (AMB). The growth of the accumulation is shown in Fig. \ref{256_len}.

We can infer that the accumulation is directly proportional to the intensity of activity or the gradient of change in activity from the above three separate cases that we have considered. Although the physics behind the three cases appear to be identical, we must include all three for comparison and the effect of varying strength of Bivariate Gaussian distribution. The following section contains a detailed examination of the growth of accumulation for the three cases. 

%LENGTH
\vspace{5cm}
\subsection{Kinetics of growing domains in high activity region}
The particles form disconnected domains in the active model B, where the activity is constant or uniform. Despite an increase in activity, the length scale for the active model remains ideal \cite{ncomms5351}. The outcomes are different for this model of inhomogeneous activity. Particles are accumulating in the middle, and the accumulation is changing as the intensity of activity increases. In the case of intensity 1, as shown in Fig 3, we see a lower accumulation than in the case of intensity 3, as shown in Fig 5. The length scale changes with a change in intensity.
We now estimate the kinetics of growing A-particle domain at the center of the box, for different intensities $G=[1.0,L/4]$, $G=[2.0,L/4]$, $G=[3.0,L/4]$. We measure the growth of A-particle domains at the center of the box in 
the following manner.
We first change the continuous density field $\phi$ to binary density $\psi=+1$ if $\phi > \phi_0$ and $\psi=-1$ if $\phi < \phi_0$, then calculate the mean size of domains $L(t)$
of A-particles in the center of the box. Mean is calculated over $20$ independent realizations. As time progresses the domain size increases $L(t)$ at the middle of the box.
We estimate the length of the growing domains at the center of the box and plot the scaled length $\frac{L(t)}{2\sigma}$ {\em vs.}time $t$ for three different intensities. 
In Fig. \ref{fig:length}, the plot is shown for 
the three cases. Very clearly growth is faster for larger intensities and saturates at late times. The $L(t)$ grows with time as, $L(t) \simeq t^{\alpha}$. We 
calculate the dynamic exponent $\alpha(t)= \frac{d \ln L(t)}{d \ln t}$. The plot of $\alpha(t)$ vs time $t$ is shown in Fig., \ref{256_beta}. For all intensities up to some intermediate time  $\alpha(t)$ remains between $2/3$ to $3/4$ and then decay to zero in the steady state. We find steady state is achieved faster for higher intensity. 
For comparison, we also calculate the characteristic length for pure passive model B and active model B with constant $\lambda$ in Fig., \ref{act_pas}. A straight line with slope 
$1/3$ is shown for the comparison of growing length for the two cases. 
We also plot the mean value of exponent $\bar{\alpha} $ for three different system sizes $L=64, 128$ and $256$ in Fig. \ref{alpha}.
The mean is obtained from the value of $\alpha(t)$ before it reaches  the steady state (approaches zero). 
Which is very different from the asymptotic steady state as found in 
kinetics of other model systems 
\cite{new_1,new_2}. 
Here due to finite system steady state is always a flat domain 
with saturated length $L(t) \approx L/2$ and $\alpha(t) \rightarrow 0$. 
Clearly the mean of the $\alpha$ for growing domain before it reaches  the 
steady state approaches value close to $0.8$ for large system sizes.
Hence, the growth of accumulation of density in middle of the box for IAMB is much faster than that for the model B
and active model B. Detail study of effect of activity on the growing domains in the active model in studied elsewere \cite{sudeep}.
%where $\alpha \sim 3/4$ to $2/3$ as we go from largest to smallest intensity in our model}. 

%\begin{table}
%\begin{center}
%\caption{ List of the values of $D_{eff}$ and $D_{est}$ for %different values of $r$ when $\alpha = 0.5$.}
%\label{table1}
%  \begin{tabular}{l*{10}{c}r}
%\hline
%{\bf$r$}              & 0.1 & 0.2 & 0.3 & 0.4 & 0.5  & 0.6 & %0.7 & 0.8 & 0.9 & 1.0 \\
%\hline
%{\bf$D_{eff}$} & 12.142  & 3.434 & 1.592 & 0.770 & 0.467  & %0.335 & 0.268 & 0.212 & 0.172 & 0.123  \\
%\hline
%{\bf$D_{est}$}           & 12.500 & 3.125 & 1.389 & 0.781 & %0.500  & 0.347 & 0.255 & 0.195 & 0.154 & 0.125 \\
%\hline
%{\bf$\beta$}           & 01.003 & 1.067 & 1.087 & 1.003 & %1.022  & 1.029 & 1.060 & 0.960 & 0.984 & 1.004  \\
%\hline
%\end{tabular}
%\end{center}
%\end{table}

%\begin{figure}%*}
%\centering
%  \includegraphics[width=10cm,height=8cm]{Deff_vs_r_alpha_05.eps}
%  \caption{(Color online) Plot for the variation of $Diffusivity$ vs. $r$ when $\alpha =0.5$. Numerical data (square) fits well with the estimated one (circle).}
%  \label{fig: 11}
%\end{figure}%*}  

\section{Discussion}
\label{discussion}
We have studied the steady state and dynamics of phase separation  of active-particles: which  experience the inhomogeneous activity on a two-dimensional
substrate. Activity parameter is distributed as Bivariate-Gaussian distribution at the center of the system, with a standard deviation of $\frac{1}{4}$*(size of the system). 
Unlike model a with constant activity parameters we have observed that accumulation of particles with variable densities are formed. 
The accumulation increases by increasing the intensity of distribution. In the region outside the distribution or away from 
the distribution, density phase separation is the same as for passive model B. \cite{chc2,chc,chc3} \\
%We had a perception that the particles would diverge away from the area of high activity(i.e, center of the system) but in contrast particles pile up in the area of high activity. This behaviour is yet to be explained statistically.From results we have noticed that the densities of accumulations changes with change in activity distribution intensity.
We also estimated the growth of A-particle domains in the center of the box with time. The domain of A-particle in the region of high activity grows as 
a power law with time. 
%The power law exponent increases with increasing intensity of activity distribution. \\

Hence our study gives an interesting steady state of density phase separation of active particles on a inhomogeneous patterned substrate and it can be useful to 
understand the trapping and transport of active particles in inhomogeneous systems. A detailed understanding of density phase separation is required to understand the mechanism 
of density phase separation in the inhomogeneous system.

%The correlation plots of different intensities were almost indistinguishable but, in systems of $X2$ and $X3$ intensities we notice small fluctuations in correlation plot. The plot of length vs time provides information of slopes($m$) or $L(t)\hspace{0.1cm} \alpha \hspace{0.1cm} t^m$. For intensity $X3$ we get a slope of $\frac{1}{3}$ or $L(t)\hspace{0.1cm} \alpha \hspace{0.1cm} t^\frac{1}{3}$, which is similar to Passive Model B. For intensity $X2$ we get a slope of $\frac{1}{4}$ or $L(t)\hspace{0.1cm} \alpha \hspace{0.1cm} t^\frac{1}{4}$. For intensity $X1$ it is variable with no constant slope. 

%\begin{acknowledgement}
%\section{Acknowledgement}
%SK and SM are thankful to the Editor for introducing equation (\ref{eq:2}) and giving useful feedback. SK and SM would like to thank DST-INSPIRE Faculty award 
%for financial support. SK would also like to thank  D. Giri,  Rajeev Singh for their useful suggestions. SM would like to thank E. G. D. Cohen for introducing the problem of the dynamics of particle on Lorentz lattice gas.
%\end{acknowledgement}
%%%%%%%%%%%%%%%%%%%%%%%%%%%%%%%%%%%%%%%%%%%%%%%%%%%%%%%%%%%%%%%%%%%%%%%%%
%End of the paper
%%%%%%%%%%%%%%%%%%%%%%%%%%%%%%%%%%%%%%%%%%%%%%%%%%%%%%%%%%%%%%%%%%%%%%%%%%%%

\end{document}